\documentclass{aa}

\usepackage{graphicx}
\usepackage{amsmath}
\usepackage{listings}
\usepackage[utf8]{inputenc}
\usepackage[T1]{fontenc}

\usepackage{lscape}
\usepackage{multirow}
\usepackage{tablefootnote}

\usepackage{tikz}
\usepackage[makeroom]{cancel}
\usetikzlibrary{shapes.geometric, arrows}
\usepackage[free-standing-units=true, range-units=single, uncertainty-mode=separate]{siunitx}
\DeclareSIUnit\mag{mag}
\DeclareSIUnit\year{yr}
\DeclareSIUnit\erg{erg}
\DeclareSIUnit\gal{galaxy}
\DeclareSIUnit\vegamag{mag_{Vega}}
\DeclareSIUnit\parsec{pc}
\DeclareSIUnit\mjd{MJD}
\DeclareSIUnit\jansky{Jy}
\DeclareSIUnit\galaxy{galaxy}

\newcommand{\redchi}{\ensuremath{\chi^2_{\mathrm{red}}}}
\newcommand{\linkorcid}[1]{\href{https://orcid.org/#1}{\includegraphics[width=8pt]{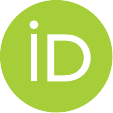}}}

\usepackage{enumitem}

\usepackage{booktabs}

\usepackage{txfonts}
\usepackage[]{hyperref}
\hypersetup{
    colorlinks=true,
    linkcolor=violet,
    filecolor=blue,
    urlcolor=cyan,
    citecolor=blue
    }

\usepackage{xcolor}
\usepackage{amsfonts}
\usepackage{color}

\graphicspath{{./figures/}}
\tikzstyle{startstop} = [rectangle, rounded corners,
minimum width=2cm,
minimum height=1cm,
text centered,
draw=black,
fill=red!30]

\tikzstyle{io} = [trapezium,
trapezium stretches=true, 
trapezium left angle=70,
trapezium right angle=110,
minimum width=2cm,
minimum height=1cm, text centered,
text width=2cm,
rounded corners=.05cm,
draw=black, fill=blue!30]

\tikzstyle{process} = [rectangle,
minimum width=2cm,
minimum height=1cm,
text centered,
text width=3cm,
draw=black,
rounded corners=.05cm,
fill=orange!30]

\tikzstyle{decision} = [diamond,
minimum width=2cm,
minimum height=1cm,
text centered,
draw=black,
text width=2cm,
rounded corners=.2cm,
fill=green!30]

\tikzstyle{arrow} = [thick,->,>=stealth,rounded corners=.1cm,]

\begin{document}

    \title{Flaires}
    \subtitle{A Comprehensive Catalog of Dust-Echo-like Infrared Flares}

    \author{
        J. Necker \inst{\ref{desy}}\fnmsep\inst{\ref{hu}}\fnmsep\thanks{jannis.necker@desy.de}\linkorcid{0000-0003-0280-7484}
        \and
        E. Graikou \inst{\ref{desy}}
        \and
        M. Kowalski \inst{\ref{desy}}\fnmsep\inst{\ref{hu}}\linkorcid{0000-0001-8594-8666}
        \and
        A. Franckowiak \inst{\ref{rub}}\linkorcid{0000-0002-5605-2219}
        \and
        J. Nordin \inst{\ref{hu}}\linkorcid{0000-0001-8342-6274}
        \and
        T. Pernice \inst{\ref{desy}}\fnmsep\inst{\ref{hu}}\linkorcid{0009-0002-1457-1712}
        \and
        S. van Velzen \inst{\ref{leiden}}\linkorcid{0000-0002-3859-8074}
        \and
        P. M. Veres \inst{\ref{rub}}\linkorcid{https://orcid.org/0000-0002-9553-2987}
      }

    \institute{
       Deutsches Elektronen Synchrotron DESY, Platanenallee 6, 15738 Zeuthen, Germany\label{desy}
       \and
       Institut für Physik, Humboldt-Universität zu Berlin, 12489 Berlin, Germany\label{hu}
       \and
       Fakultät für Physik \& Astronomie, Ruhr-Universität Bochum, 44780 Bochum, Germany\label{rub}
       \and
       Leiden Observatory, Leiden University, Postbus 9513, 2300 RA, Leiden, Netherlands\label{leiden}
    }

    \date{Received; accepted }

    \abstract
    {Observations of transient emission from extreme accretion events onto supermassive black holes can reveal conditions in the center of galaxies and the black hole itself. Most recently, they have been suggested to be emitters of high-energy neutrinos. If it is suddenly rejuvenated accretion or a tidal disruption event (TDE) is not clear in most cases.}
    {We expanded on existing samples of infrared flares to compile the largest and most complete list available. A large sample size is necessary to provide high enough statistics for far away and faint objects to estimate their rate. Our catalog is large enough to facilitate a preliminary study of the rate evolution with redshift for the first time.}
    {We compiled a sample of 40 million galaxies, and, using a custom, publicly available pipeline, analyzed the WISE light curves for these 40 million objects using the Bayesian Blocks algorithm. We selected promising candidates for dust echos of transient accretion events and inferred the luminosity, extension, and temperature of the hot dust by fitting a blackbody spectrum.}
    {We established a clean sample of 823 dust-echo-like infrared flares of which we can estimate the dust properties for 568. After removing 70 objects with possible contribution by synchrotron emission, the luminosity, extension, and temperature are consistent with dust echos. Estimating the dust extension from the light curve shape revealed that the duration of the incident flare is broadly compatible with the duration of TDEs. The resulting rate per galaxy is consistent with the latest measurements of infrared-detected TDEs and appears to decline at increasing redshift.}
    {Although systematic uncertainties may impact the calculation of the rate evolution, this catalog will enable further research in phenomena related to dust-echos from TDEs and extreme accretion flares.}


    \maketitle
    %

    \section{Introduction}
    \label{sec:intro}

    Accretion onto compact objects is the most efficient form of energy release of normal matter. The heaviest such objects are the Supermassive Black Holes (SMBH) that reside in the center of most galaxies \citep{magorrianDemographyMassiveDark1998}. If matter is accreted onto the SMBH it forms a luminous accretion disk, dissipating the gravitational energy throughout the electromagnetic spectrum \citep{lynden-bellGalacticNucleiCollapsed1969, reesBlackHoleModels1984}. \\
    If a star passes too close to the SMBH, the tidal forces can rip it apart. Its debris's subsequent accretion is visible as a luminous transient dubbed a Tidal Disruption Event (TDE) \citep{reesTidalDisruptionStars1988}. TDEs can last months up to years and offer unique insights into the properties of previously inactive SMBHs \citep{kesdenTidaldisruptionRateStars2012, mocklerWeighingBlackHoles2019, wenLibrarySyntheticXRay2022}. They were originally discovered as X-ray transients with a soft, thermal spectrum \citep{komossaDiscoveryGiantLuminous1999}. However, the optical detection rate dramatically increased in recent years thanks to the advent of large time-domain survey instruments, and TDEs are now frequently discovered by instruments like the Zwicky Transient Facility (ZTF; \citealt{bellmZwickyTransientFacility2018, velzenSeventeenTidalDisruption2021}), the All-Sky Automated Survey for SuperNovae (ASAS-SN; \citealt{kochanekAllSkyAutomatedSurvey2017}, e.g. \citealt{holoienASASSN14aeTidalDisruption2014, hinkleDiscoveryFollowupASASSN19dj2021}) and the Panoramic Survey Telescope and Rapid Response System (Pan-STARRS; \citealt{Chambers:2016aa}, e.g. \citealt{gezariUltravioletOpticalFlare2012, chornockULTRAVIOLETBRIGHTSLOWLYDECLINING2013, nichollTidalDisruptionEvent2019}). Indeed, the optical regime became the dominating detection channel with tens of TDEs per year \citep{gezariTidalDisruptionEvents2021}. In the future, this is expected to increase with the Legacy Survey of Space and Time (LSST) of the Vera Rubin Observatory to thousands per year \citep{ivezicLSSTScienceDrivers2019}. However, the bulk of the emission is expected in the extreme-UV which is almost impossible to observe because of absorption by neutral hydrogen \citep{ulmerFlaresTidalDisruption1999, luInfraredEmissionTidal2016}. Consequently, TDEs have been detected in the UV \citep{gezariUltravioletDetectionTidal2006} with an expected increase to hundreds per year with the survey of ULTRASAT \citep{sagivSCIENCEWIDEFIELDUV2014}. But the radiation energy released by TDEs can be absorbed by dust in the vicinity of the SMBH and re-emitted in the infrared with the peak of the emission between \qtyrange{3}{10}{\micro\meter} and a typical luminosity of \qtyrange{e42}{e43}{\erg\per\second} \citep{luInfraredEmissionTidal2016}. This infrared emission has been observed in follow-up studies of UV/O TDEs \citep{velzenDISCOVERYTRANSIENTINFRARED2016, jiangWISEDetectionInfrared2016, douLONGFADINGMIDINFRARED2016}. The transient infrared emission has also been used to identify TDE candidates \citep{wangLongtermDeclineMidinfrared2018}. The most notable effort is the TDE sample compiled based on the WISE transient pipeline (WTP), a list based on photometry, geared towards high purity \citep{mastersonNewPopulationMidinfraredselected2024}. This revealed that the preference for post-starburst host galaxies, possibly due to temporarily enhanced rates as a consequence of a galaxy merger \citep{hammersteinTidalDisruptionEvent2021} is weaker when adding IR selected to optically detected TDE candidates. It also provides a rate measurement of $\qty{2(.3)e-5}{\per\galaxy\per\year}$ in consistency with recent measurements in the optical \citep{yaoTidalDisruptionEvent2023}. Where the first estimates of the rate were almost an order of magnitude lower \citep{donleyLargeAmplitudeXRayOutbursts2002} than the theoretically expected rate of around \qty{e-4}{\per\galaxy\per\year} there is now only a factor of four to five discrepancy. It suggests that the infrared domain opens a detection channel for the same intrinsic events just with a higher dust covering factor, the percentage of the emission absorbed by dust.
    \\
    If the accretion onto the SMBH is continuous, we observe the object as an Active Galactic Nucleus (AGN) which is the most luminous steady source in the universe.  Although the accretion is roughly constant, random fluctuation is inherent to AGN activity. These AGN flares can happen on a timescale from hours to years \citep{ulrichVARIABILITYACTIVEGALACTIC1997} and show brightness variations of the order of tens of percent \citep{berkEnsemblePhotometricVariability2004}.  Apart from this, there is an increasing number of observations of AGN variability beyond the expected statistical fluctuations. The first class is a long-term decline or rise over several years together with a change in the spectral features \citep{lamassaDISCOVERYFIRSTCHANGING2015, gezariIPTFDiscoveryRapid2017, frederickNewClassChanginglook2019, yangProbingOriginChanginglook2023} called changing-look AGN. These are typically attributed to accretion rate changes \citep{husemannCloseAGNReference2016a, yangProbingOriginChanginglook2023}. In contrast to this, there is a class of major AGN flares that are characterized by a sudden increase in their UVO emission that differentiates them from the changing-look phenomena and are attributed to a sudden rejuvenation of the accretion onto the SMBH \citep{grahamUnderstandingExtremeQuasar2017, trakhtenbrotNewClassFlares2019}. \\
    Similar to the dust echoes of TDEs, the dust around the SMBH can reprocess the electromagnetic emission and re-emit it in the infrared \citep{barvainisHotDustNearInfrared1987}. Because the dust efficiently absorbs from the optical to X-ray wavelengths, the resulting dust echo is agnostic to the nature of the incident transient as long as the transient evolution timescale is smaller than the light-travel time to distances where the dust is stable. Therefore infrared transient emission can be used to compile a sample of TDEs and extreme accretion events at the same time. Indeed, \citet{jiangMidinfraredOutburstsNearby2021} recently compiled a sample of Mid-infrared Outbursts in Nearby Galaxies (MIRONG), a similar ensemble of infrared flares, concluding that these flares are likely due to extreme transient AGN accretion events or TDEs. \\
    Transient accretion events of SMBHs have been suggested as the sources of high-energy particles \citep{farrarGIANTAGNFLARES2009}. This is supported in particular by the detection by the IceCube neutrino observatory of the flaring blazar TXS 0506+056 \citep{icecubecollaborationMultimessengerObservationsFlaring2018} in coincidence with a high-energy neutrino alert \citep{aartsenIceCubeRealtimeAlert2017}. These alerts are sent out by IceCube in real time and followed up by many observatories \citep{neckerASASSNFollowupIceCube2022, steinNeutrinoFollowupZwicky2023}. ZTF detected two transients connected to SMBH accretion coincident with two high-energy neutrino alerts: AT2019dsg \citep{steinTidalDisruptionEvent2021}, a spectroscopically classified TDE, and AT2019fdr \citep{reuschCandidateTidalDisruption2022}, a giant AGN flare (see \citet{pitikHighenergyNeutrinoEvent2022} for an alternative interpretation). The unifying feature is a large dust echo. A systematically compiled sample of 63 similar flares based on optical ZTF detections revealed a third event: AT2019aalc, another giant AGN flare. The sample is correlated with IceCube's high-energy neutrino alerts at $3.6\sigma$ \citep{vanvelzenEstablishingAccretionFlares2024} which suggests a significant contribution from these accretion events to the diffuse neutrino flux \citep{icecubecollaborationEvidenceHighEnergyExtraterrestrial2013}. Although an additional search using the full neutrino data sample did not show any significant excess \citep{neckerSearchHighEnergyNeutrinos2023}, the size of the accretion flare sample was small and restricted to flares with an optical counterpart detection by ZTF which started observing in 2018. To harvest the full potential of the neutrino data, a sample that spans as much of the IceCube data-taking period since 2010, does not rely on optical detections, and spans as much of the sky as possible is needed. Additionally, because neutrinos are not attenuated traveling through the universe, this sample should be as complete as possible. \\
    In this paper, we will present a catalog of infrared flares detected by WISE that expands on existing samples by an order of magnitude. This catalog provides a unique opportunity to tackle both the rate evolution and the neutrino association because it covers the whole sky, extends to high redshifts, and goes back to the start of IceCube operations.
    \\
    The paper is organized as follows: In Section \ref{sec:data} we detail the creation of the infrared light curves and the selection of dust-echo-like flares. In Section \ref{sec:res} we explore the most likely physical origin of these flares and evaluate the corresponding completeness of the resulting catalog in Section \ref{sec:completeness}. We investigate the implications for the rate evolution in Section \ref{sec:rate} before discussing our results in Section \ref{sec:discussion}. \\
    Throughout this paper, we assume a flat $\Lambda$CDM cosmological model with the parameters presented in \citet{planckcollaborationPlanck2018Results2020}.


    \section{Data}
    \label{sec:data}

    WISE was launched at the end of 2009 and began surveying the sky in four bands at the beginning of 2010 until it was shut down after its nominal mission lifetime in 2011. The data products of this phase were combined under the name AllWISE. WISE was re-activated in 2013 with the primary science goal of discovering near-Earth objects \citep{mainzerINITIALPERFORMANCENEOWISE2014}, which is called the NEOWISE reactivation phase (NEOWISE-R). Since then it has continuously observed the sky in two bands at \qty{3.4}{\micro\meter} and \qty{4.6}{\micro\meter} (W1 and W2). In this work, we use AllWISE and NEOWISE-R data to identify dust echoes of possible extreme accretion events because the two filters cover the relevant wavelength range, the sampling rate of around two data points per year is sufficient and the observations span around ten years.


    \subsection{Parent Galaxy Sample}
    \label{sec:parent_sample}
    To know where in the sky to look for candidate IR flares, we need the positions of galaxies, called our parent galaxy sample. This should be as complete as possible with as much sky coverage as possible. The NEWS sample \citep{khramtsovNorthernExtragalacticWISE2020} provided the biggest such catalog at the start of our analysis with around $40 \times 10^6$ extragalactic objects identified by a machine learning algorithm based on WISE measurements from the AllWISE source catalog and PSF magnitude measurements from the Pan-STARRS Data Release 1 catalog. The benefit is that each source has an associated identifier in the AllWISE source catalog which allows us to easily query the WISE database for the corresponding single-exposure measurements (see Section \ref{sec:timewise}). To reject bright foreground stars and achieve high purity, the NEWS sample excludes sources brighter than $\qty{14}{\mag}$ in $g$ and $r$-band. \\
    To include the bright sources, we used the WISE-PS1-STRM sample \citep{beckWISEPS1STRMNeuralNetwork2022}. It also uses a neural network and WISE and Pan-STARRS photometry to infer object classifications and photometric redshifts. We selected probable galaxies ($\mathtt{prob\_Galaxy} > 0.5$) with a $g$-band magnitude brighter than $\qty{20}{\mag}$ ($\mathtt{gFPSFMag} \leq 20$). We only used non-extrapolated classifications, so objects that are within the bounds of the training set ($\mathtt{extrapolation\_Class} = 0$). We crossmatched the resulting $3.2 \times 10^6$ objects to the NEWS sample based on the Pan-STARRS object identifier and found $9.6 \times 10^5$ non-overlapping sources. \\
    To ensure we are as complete as possible in the local universe, we included the Local Volume Sample from the NASA Extragalactic Database (NED-LVS; \citealt{cookCompletenessNASAIPAC2023}). It contains around $1.8 \times 10^6$ sources with a redshift up to $z = 0.2$. We include $6.7 \times 10^5$ sources from  NED-LVS that have no counterpart in our sample within a radius of $1 \arcsec$. \\
    The resulting parent galaxy sample has $4.2 \times 10^7$ sources. The majority is located in the northern 3/4 of the sky above $\delta \gtrsim \qty{-30}{\degree}$ since both the NEWS and the WISE-PS1-STRM samples are based on Pan-STARRS data. All sources with $\delta \lesssim \qty{-30}{\degree}$ come from the NED-LVS. \\
    For the around $1.9 \times 10^6$ sources in NED-LVS, we take the redshift from that catalog and for another $3.2 \times 10^7$ sources we take the photometric redshift estimate from WISE-PS1-STRM. In total, we have at least photometric redshift estimates for $3.4 \times 10^7$ sources, around 80\% of all sources in our parent sample.


    \subsection{Infrared Light Curves}
    \label{sec:timewise}
    WISE scans the sky continuously, taking \qty{7.7}{\second} exposures. Roughly every half year each point in the sky will be covered by the FoV for about 12 consecutive orbits for about a day. We used the PSF fit photometry measurements based on these single exposures from both the AllWISE and the NEOWISE-R periods to get information about the evolution of the infrared luminosity over time. The notable difference is that the AllWISE photometry is measured at the position of the corresponding source in the AllWISE source catalog while the NEOWISE-R data fits the position of the source and only references the associated AllWISE catalog source.\\
    Because photometry measurements can be impacted by image artifacts, cosmic rays, blending of close sources, and stray light from the moon, it is important to select reliable data points. We required good image quality, no de-blending, no contamination by image artifacts, a separation to the Southern Atlantic Anomaly of more than \qty{5}{\degree}, and a position outside the moon mask. \\
    The selection of data for each source depends on the origin of the individual source.
    For all sources in the NEWS sample, we got the AllWISE source catalog identifier based on the designation. We then downloaded all AllWISE multi-epoch photometry for this identifier and all associated NEOWISE-R single-exposure photometry. The sources from the WISE-PS1-STRM and the NED-LVS samples do not have an associated entry in the AllWISE database, so instead, we selected the data based on the position. The FWHM of the WISE PSF is around \qty{6}{arcsec} so we downloaded all the single exposure photometry (both AllWISE and NEOWISE-R) within that radius. This will include all data for extended or faint objects where the scatter in the best-fit position is large. However, it also poses the risk of including data points related to a neighboring object. Especially the AllWISE data is prone to this mistake because it includes the single epoch measurements or upper limits also of faint objects that are only detected in the AllWISE source catalog. Because in these cases blending will be an issue, the AllWISE photometry of an object might not pass the quality criteria mentioned above. So it is not sufficient to simply select the closest AllWISE photometry. Instead, we address the issue by finding clusters of data points and associating the closest cluster to the corresponding source in the parent sample. We employed the clustering algorithm HDBSCAN \citep{mcinnesAcceleratedHierarchicalDensity2017} to find the clusters of data points. We selected all points that belong to the closest cluster if that cluster is within \qty{1}{\arcsec}. Additionally, we selected all datapoints within \qty{1}{\arcsec}, even if they belong to a different or no cluster at all. This allows us to select the corresponding single epoch photometry also where the positional scatter is large while minimizing the risk of including unrelated data points.\\
    We stacked the single epoch photometry for each visit to get a more robust measurement. To combine the counts measured at the instrument level, we calculated the photometric zero-point $ZP_{i,j}$ for each single exposure measurement $i$ of visit $j$ per band:
    \begin{equation}
        ZP_{\nu, i,j} = V_{i,j} + 2.5 \log_{10}(C_{i,j})
    \end{equation}
    where $V_{i,j}$ is the apparent magnitude in the Vega system and $C_{i,j}$ is the flux in instrument counts. For each visit $j$, we calculated the median of the zero-point $ZP_{\nu, \mathrm{med}, j}$ per band. The spectral flux densities for each exposure are then:
    \begin{equation}
        F_{\nu, i, j} = C_{i,j} \cdot  F_{\nu, 0} \cdot \log_{10}(ZP_{\nu, \mathrm{med}, j} / 2.5)
    \end{equation}
    where $F_{\nu, 0}$ is the zero magnitude flux density \citep{jarrettSPITZERWISESURVEY2011}. Finally, the stacked flux density is the median of the spectral flux densities per visit and band. The uncertainty is the maximum of the standard deviation to the median and the Pythagorean sum of the individual measurement uncertainties. We noted outliers with unreasonable high fluxes, probably caused by cosmic rays or spurious detections that escaped the WISE flagging or asteroids that are passing through the line of sight. We found that we can reject these if they deviate from the median more than 20 times the 70th percentile. \\
    For each light curve and band, we calculated the reduced chi-square $\redchi$ with respect to the median as a preliminary measure of the variability of the light curve. This allows us to identify quiescent light curves early in the analysis based on a low $\redchi$ value. To establish a meaningful threshold, we ran the flare search algorithm as described in the next Section for a sub-sample of $2\times10^6$ light curves. It turned out that all light curves that show flaring have a value of $\redchi > 1$ in both bands. Around half of the sources do not pass this threshold and were not analyzed further (see Figure \ref{fig:flowchart}). \\
    The efficient download, stacking, and calculation of $\redchi$ as a preliminary variability metric as described in this section is implemented in the publicly available python package \texttt{timewise} \footnote{\url{https://timewise.readthedocs.io/en/latest/}} \citep{neckerJannisNeTimewiseV02024}.


    \subsection{Flare Selection}
    \label{sec:sup}

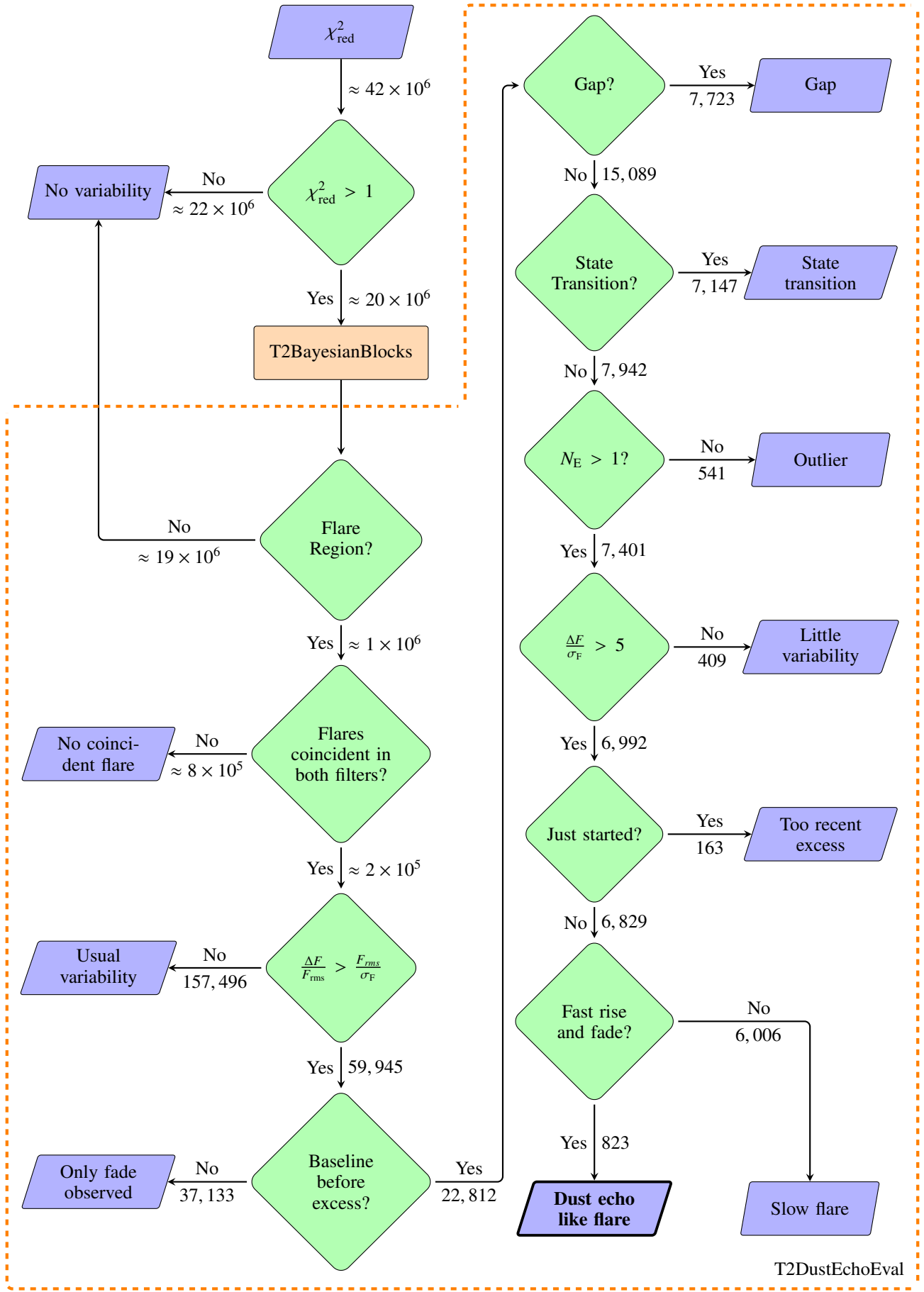
\begin{figure*}
        \centering
        \begin{center}

    \begin{tikzpicture}[node distance=2cm]

        \node (chi2_in) [io] {$\chi^2_\mathrm{red}$};
        \node (chi2_dec) [decision, below of=chi2_in, yshift=-1cm] {$\chi^2_\mathrm{red} > 1$};
        \node (baseline) [io, left of=chi2_dec, xshift=-2.5cm] {No variability};

        \node (bb) [process, below of=chi2_dec, yshift=-1cm] {\lstinline{T2BayesianBlocks}};

        \node (excess_dec) [decision, below of=bb, yshift=-1.5cm] {Flare Region?};

        \node (coinc_dec) [decision, below of=excess_dec, yshift=-2cm] {Flares coincident in both filters? };
        \node (coinc) [io, left of=coinc_dec, xshift=-2.5cm] {No coincident flare};

        \node (strength_dec) [decision, below of=coinc_dec, yshift=-2cm] {$\frac{\Delta F}{F_{\mathrm{rms}}} > \frac{F_{rms}}{\sigma_\mathrm{F}}$};
        \node (little_strength) [io, left of=strength_dec, xshift=-2.5cm] {Usual variability};

        \node (only_fading_dec) [decision, below of=strength_dec, yshift=-2cm] {Baseline before excess?};
        \node (only_fading) [io, left of=only_fading_dec, xshift=-2.5cm] {Only fade observed};

        \node (gap_dec) [decision, right of=chi2_dec, xshift=2.7cm, yshift=2cm] {Gap?};
        \node (gap) [io, right of=gap_dec, xshift=2.2cm] {Gap};

        \node (stage_transition_dec) [decision, below of=gap_dec, yshift=-1.5cm] {State Transition?};
        \node (stage_transition) [io, right of=stage_transition_dec, xshift=2.2cm] {State transition};

        \node (outlier_dec) [decision, below of=stage_transition_dec, yshift=-1.5cm] {$N_\mathrm{E} > 1$?};
        \node (outlier) [io, right of=outlier_dec, xshift=2.2cm] {Outlier};

        \node (excess_sig_dec) [decision, below of=outlier_dec, yshift=-1.5cm] {$\frac{\Delta F}{\sigma_\mathrm{F}} > 5$};
        \node (excess_sig) [io, right of=excess_sig_dec, xshift=2.2cm] {Little variability};

        \node (starting_dec) [decision, below of=excess_sig_dec, yshift=-1.5cm] {Just started?};
        \node (starting) [io, right of=starting_dec, xshift=2.2cm] {Too recent excess};

        \node (times_dec) [decision, below of=starting_dec, yshift=-1.5cm] {Fast rise and fade?};

        \node (class1) [io, below of=times_dec, yshift=-1.5cm, line width=0.5mm] {\textbf{Dust echo like flare}};
        \node (class2) [io, right of=class1, xshift=2cm] {Slow flare};

        \draw[arrow] (chi2_in) -- node[anchor=west]{$\approx 42 \times 10^6$}(chi2_dec);
        \draw[arrow] (chi2_dec) -- node[anchor=south]{No} node[anchor=north]{$\approx 22 \times 10^6$}(baseline);
        \draw[arrow] (chi2_dec) -- node[anchor=east]{Yes} node[anchor=west]{$\approx 20\times 10^{6}$}(bb);
        \draw[arrow] (bb) -- (excess_dec);
        \draw [arrow] (excess_dec) -| node[anchor=south, xshift=1.5cm]{No} node[anchor=north, xshift=1.5cm]{$\approx 19 \times 10^{6}$}(baseline);
        \draw [arrow] (excess_dec) -- node[anchor=east]{Yes} node[anchor=west]{$\approx 1 \times 10^{6}$} (coinc_dec);
        \draw[arrow](coinc_dec) -- node[anchor=north]{$\approx 8\times 10^5$} node[anchor=south]{No}(coinc);
        \draw[arrow](coinc_dec) -- node[anchor=west]{$\approx 2\times10^5$} node[anchor=east]{Yes}(strength_dec);
        \draw[arrow](strength_dec) -- node[anchor=south]{No} node[anchor=north]{$157,496$} (little_strength);
        \draw[arrow](strength_dec) -- node[anchor=west]{$59,945$} node[anchor=east]{Yes}(only_fading_dec);
        \draw[arrow](only_fading_dec) -- node[anchor=south]{No} node[anchor=north]{$37,133$}(only_fading);
        \draw[arrow](only_fading_dec) -- node[anchor=north]{$22,812$} node[anchor=south]{Yes} +(3cm, 0) |- (gap_dec);
        \draw[arrow](gap_dec) -- node[anchor=south]{Yes} node[anchor=north]{$7,723$} (gap);
        \draw[arrow](gap_dec) -- node[anchor=east]{No} node[anchor=west]{$15,089$} (stage_transition_dec);
        \draw[arrow](stage_transition_dec) -- node[anchor=south]{Yes} node[anchor=north]{$7,147$}(stage_transition);
        \draw[arrow](stage_transition_dec) -- node[anchor=east]{No} node[anchor=west]{$7,942$}(outlier_dec);
        \draw[arrow](outlier_dec) -- node[anchor=south]{No} node[anchor=north]{$541$}(outlier);
        \draw[arrow](outlier_dec) -- node[anchor=west]{$7,401$} node[anchor=east]{Yes}(excess_sig_dec);
        \draw[arrow](excess_sig_dec) -- node[anchor=south]{No} node[anchor=north]{$409$}(excess_sig);
        \draw[arrow](excess_sig_dec) -- node[anchor=east]{Yes} node[anchor=west]{$6,992$} (starting_dec);
        \draw[arrow](starting_dec) -- node[anchor=south]{Yes} node[anchor=north]{$163$} (starting);
        \draw[arrow](starting_dec) -- node[anchor=west]{$6,829$} node[anchor=east]{No}(times_dec);
        \draw[arrow](times_dec) -- node[anchor=east]{Yes} node[anchor=west]{$823$}(class1);
        \draw[arrow](times_dec) -| node[anchor=south east, xshift=-0.6cm]{No} node[anchor=north east, xshift=-0.4cm]{$6,006$}(class2);

        \node (box1) [below of=bb, xshift=-6.2cm, yshift=1cm]{};
        \node (box2) [right of=box1, xshift=6.5cm]{};
        \node (box3) [above of=gap, xshift=1.8cm, yshift=-0.5cm]{};
        \node (box4) [below of=box3, yshift=-22cm]{};
        \node[above left] at (box4.north west){\lstinline{T2DustEchoEval}};
        \draw[ultra thick, dashed, rounded corners=.1cm, orange] (box1) -- (box2) |- (box3) -- (box4) -| (box1);

    \end{tikzpicture}

\end{center}
        \caption{Flowchart of the flare selection process. See Section \ref{sec:sup} for a detailed explanation of the selection steps.}
        \label{fig:flowchart}
    \end{figure*}

    To find any potential excess in the light curves we use the Bayesian Blocks algorithm \citep{scargleSTUDIESASTRONOMICALTIME2013} as implemented in \texttt{astropy} \citep{theastropycollaborationAstropyProjectSustaining2022}. It divides the light curve into intervals of constant flux. Fewer intervals are favored where the strength of the preference depends on a prior. We chose the empirically motivated prior for point measures for the number of change points
    \begin{equation}
        N_{\mathrm{cp,\, prior}} = 1.32 + 0.577 \, \log_{10}(N_\mathrm{visit})
    \end{equation}
    for a light curve with $N_\mathrm{visit}$ visits \citep{scargleSTUDIESASTRONOMICALTIME2013}. We defined the quiescent state, i.e. baseline, as the lowest flux we measured across the light curve. This does not have to be one continuous interval but can be interrupted by excesses. To combine all baseline intervals into one measurement, we successively combined the blocks with the lowest spectral flux density with the baseline. The updated baseline is the mean of all data points in the baseline blocks and its uncertainty is the standard deviation. We stopped when there was no more block within $5 \sigma$ of the baseline. All blocks not within the baseline are considered excess blocks. Adjacent excess blocks are considered as one excess and its start and end are defined as the time of the first and last data point within the excess, respectively. \\
    To efficiently analyze the $\mathcal{O}(10^6)$ light curves, we implemented the procedure in the \texttt{ampel} framework \citep{nordinTransientProcessingAnalysis2019}, which provides high-throughput streaming of data and is capable of including user-contributed code in the form of so-called units. \texttt{ampel} pipelines are split into four stages called tiers. The Bayesian Blocks analysis is implemented in tier two as the unit \texttt{T2BayesianBlocks} (see Figure \ref{fig:flowchart}).

    The Bayesian Blocks result lets us identify flaring behavior. In the next step, we aim to quantify the significance of the flare compared to the rest of the light curve to identify significant dust-echo-like flares. We implemented the following cuts as another tier two unit \texttt{T2DustEchoEval} (see Figure \ref{fig:flowchart} for the corresponding numbers):
    \begin{enumerate}
        \item \textit{Flare Region}: an excess in at least one of the two WISE filters
        \label{sup:excess}
        \item \textit{Flares coincident in both filters}: a coincident excess in the other band
        \label{sup:coincidence}
        \item $\mathit{\Delta F / F_\mathrm{rms} > F_\mathrm{rms} / \sigma_\mathrm{F}}$: dust echo strength $\Delta F / F_\mathrm{rms}$ higher than the significance of variability in the extraneous part of the light curve $F_\mathrm{rms} / \sigma_\mathrm{F}$ \citep{vanvelzenEstablishingAccretionFlares2024} to make the distinction to stochastic AGN variability. $\Delta F$ is the difference between baseline and peak flux, $F_\mathrm{rms}$ the root-mean-square of the extraneous light curve and $\sigma_\mathrm{F}$ the standard deviation of the baseline.
        \label{sup:strength}
        \item \textit{Baseline before excess}: at least one baseline datapoint before the excess
        \label{sup:baseline}
        \item \textit{Gap}: a baseline datapoint not more than \qty{500}{\day} before the first detection of the excess. This criterion excludes flares that rise during the \textit{AllWISE} period.
        \label{sup:gap}
        \item \textit{State transition}: a segmentation of the excess into at least two blocks to reject state transitions or a difference in time between the first and last datapoint in the excess of no more than \qty{3}{\year}
        \item \textit{$N_\mathrm{E} > 1$}: more than one datapoint in the excess
        \label{sup:outlier}
        \item \textit{$\Delta F / \sigma_\mathrm{F} > 5$}: a detection significance of the excess of at least $5 \sigma$ with respect to the baseline.
        \item \textit{Just started}: If the excess consists of a single block at the end of the light curve we consider it too recent.
        \label{sup:just_started}
    \end{enumerate}
    At this stage, we calculated the e-folding rise and fade times as
    \begin{equation}
        \tau = \frac{\Delta t}{\log(F_2/F_1)}.
    \end{equation}
    For the rise time, $\Delta t$ is the time between the last baseline datapoint and the maximum of the light curve, $F_1$ is the baseline spectral flux density, and $F_2$ is the peak spectral flux density. For the fade time, $\Delta t$ is the time from maximum to the last datapoint in the excess, $F_1$ is the peak spectral flux density, and $F_2$ is the last measurement of the excess. We make the distinction between dust echo-like flares and flares on longer timescales based on these e-folding times:
    \begin{enumerate}[resume]
        \item \textit{Fast rise and fade?}: $\tau_\mathrm{rise} < \qty{1000}{\day}$ and $\qty{400}{\day} < \tau_\mathrm{fade} < \qty{5000}{\day}$.
    \end{enumerate}
    After these cuts, we identified 823 flares that meet our criteria for dust-echo-like infrared flares. \\
    This flare selection pipeline is publicly available as the \textit{\texttt{timewise} Subtraction Pipeline} python package, \texttt{timewise-sup}\footnote{\url{https://jannisnecker.pages.desy.de/timewise_sup/docs}} \citep{neckerTimewisesupTimewiseSubtraction2024}.

    \subsection{Difference Light Curves}
    To analyze the flare properties, we want to obtain light curves that only contain the contribution from the excess. To get this difference photometry, we subtracted the baseline, measured as described in the previous section from the data (see Section \ref{sec:timewise}). The uncertainty is the Pythagorean sum of the baseline uncertainty and the uncertainty of the stacked data. The spectral flux $\nu F_\nu$ is then the difference spectral flux density $F_\nu$ multiplied by the frequency $\nu$ corresponding to the central wavelength of the WISE bandpass.
    \\
    A caveat of using the single exposure photometry is that there is a bias for dim sources near or below the limiting magnitude of $m_\mathrm{W1} \approx 16.0$ and $m_\mathrm{W2} \approx 14.9$. The scatter due to instrument noise in single-exposure photometry still produces detections above these limits, biasing the baselines toward brighter values. This is known as the Eddington bias \citep{eddingtonFormulaCorrectingStatistics1913} and has been identified as an effect already for the Infrared Astronomical Satellite \citep{beichmanInfraredAstronomicalSatellite1988} as well as for NEOWISE-R \citep{mainzerINITIALPERFORMANCENEOWISE2014}. Because the bias gets stronger the dimmer the sources, this will mainly impact the baseline measurements. We verified that the conclusions of this analysis do not change when replacing our baseline measurement with the photometry values from the deeper AllWISE source catalog. We still used the measured values for the difference photometry to ensure we found the lowest state even if there is some excess in the AllWISE period.
    \\
    To obtain redshifts, we used the well-curated values from the NED-LVS, which are available for 135 sources. If they are not available, we used the photometric redshifts from WISE-PS1-STRM (411 sources). For 139 sources we found redshifts in other catalogs (see Table \ref{tab:redshift_cats}). Where we found more than one match, we divided the matches into categories based on spectroscopic and photometric measurements and the match distance and took the mean of the best category. We calculated the luminosity per bandpass in the rest frame based on the luminosity distance
    \begin{equation}
        \nu_\mathrm{e}  L_{\nu_\mathrm{e}} = \nu  F_{\nu} \cdot  4 \pi d_L^2(z)
    \end{equation}
    where the subscript $\mathrm{e}$ indicates the rest frame quantities.

    \begin{table}
        \begin{tabular}{rcc}
            \toprule
            \textbf{Catalog} & \textbf{matches} & \textbf{Reference} \\
            \midrule
            WISE x SuperCOSMOS & 66 & \citet{bilickiWISESuperCOSMOSPHOTOMETRIC2016} \\
            SDSS DR10 & 5 & \citet{bresciaAutomatedPhysicalClassification2015} \\
            GLADE & 7 & \citet{dalyaGLADEGalaxyCatalogue2018} \\
            LS DR8 & 75 & \citet{duncanAllpurposeAllskyPhotometric2022} \\
            NEDz & 29 & \tablefootnote{\url{https://ned.ipac.caltech.edu/}} \\
            \bottomrule
        \end{tabular}
        \caption{Number of matches in redshift catalogs}
        \label{tab:redshift_cats}
    \end{table}


    \section{Physical Origin}
    \label{sec:res}


    \subsection{Dust Echo Scenario}
    \label{sec:bb}

    So far we established a sample of 823 objects showing an IR flare that is significant compared to the extraneous part of the light curve. Although we have already implicitly assumed that these IR excesses are dust echoes from nuclear flares in the selection process in Section \ref{sec:sup}, we analyze the physical properties corresponding to the dust echo scenario in more detail to check for consistency.\\

\begin{figure*}
        \centering
        \includegraphics[width=0.99\textwidth]{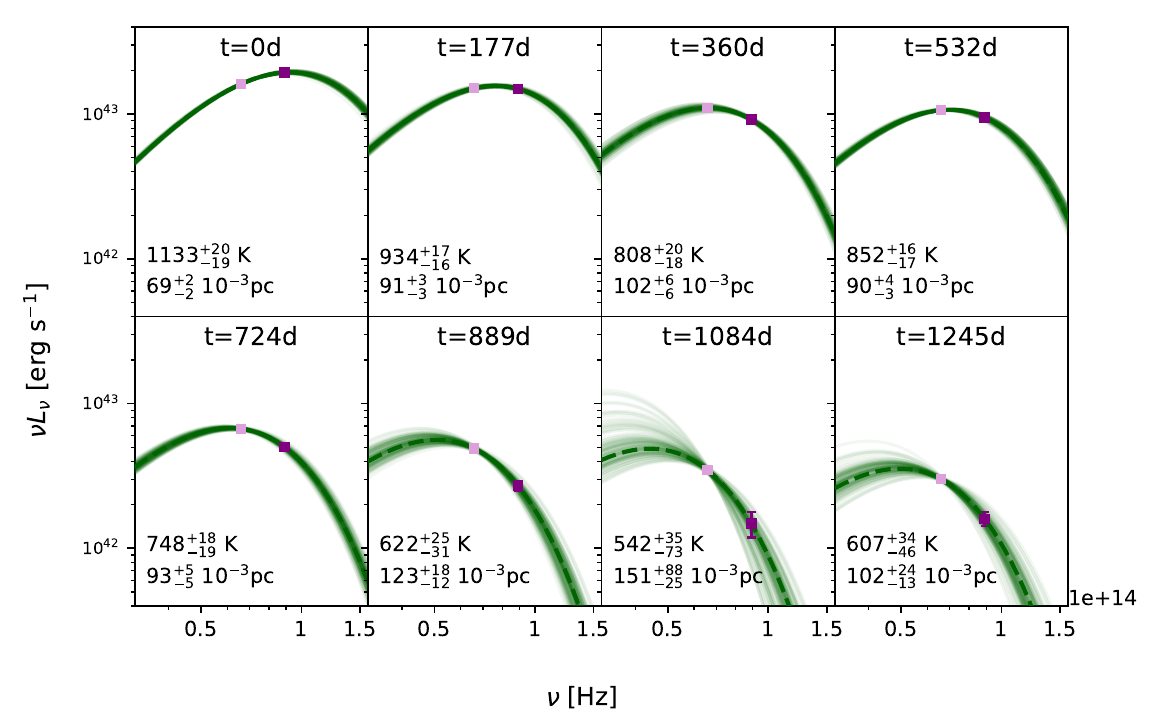}
        \caption{Blackbody fits for NGC 7392 for all epochs of the flare where the MCMC sampling resulted in meaningful constraints on the fit parameters. The dashed green line shows the best fit and the transparent lines the results from the MCMC sampling. The best fit and 68th percentile temperature and effective radius values are shown in the bottom left corner. The frequency $\nu$ is given in the source frame.}
        \label{fig:ngc7392_bb}
    \end{figure*}
\begin{figure}
        \centering
        \includegraphics[width=0.49\textwidth]{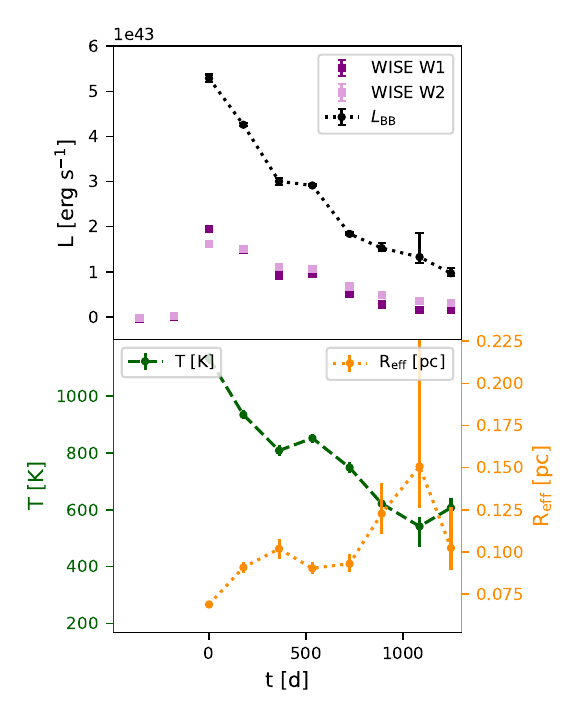}
        \caption{Light curve and evolution of $R_\mathrm{eff}$ and $T$ for NGC 7392. The crosses mark the best-fit values of the visit where the MCMC sampling did not result in a meaningful constraint of $L_\mathrm{bol}$. The results are consistent with \citet{panagiotouLuminousDustobscuredTidal2023a}.}
        \label{fig:ngc7392_lc}
    \end{figure}
\begin{figure}
        \centering
        \includegraphics[width=.49\textwidth]{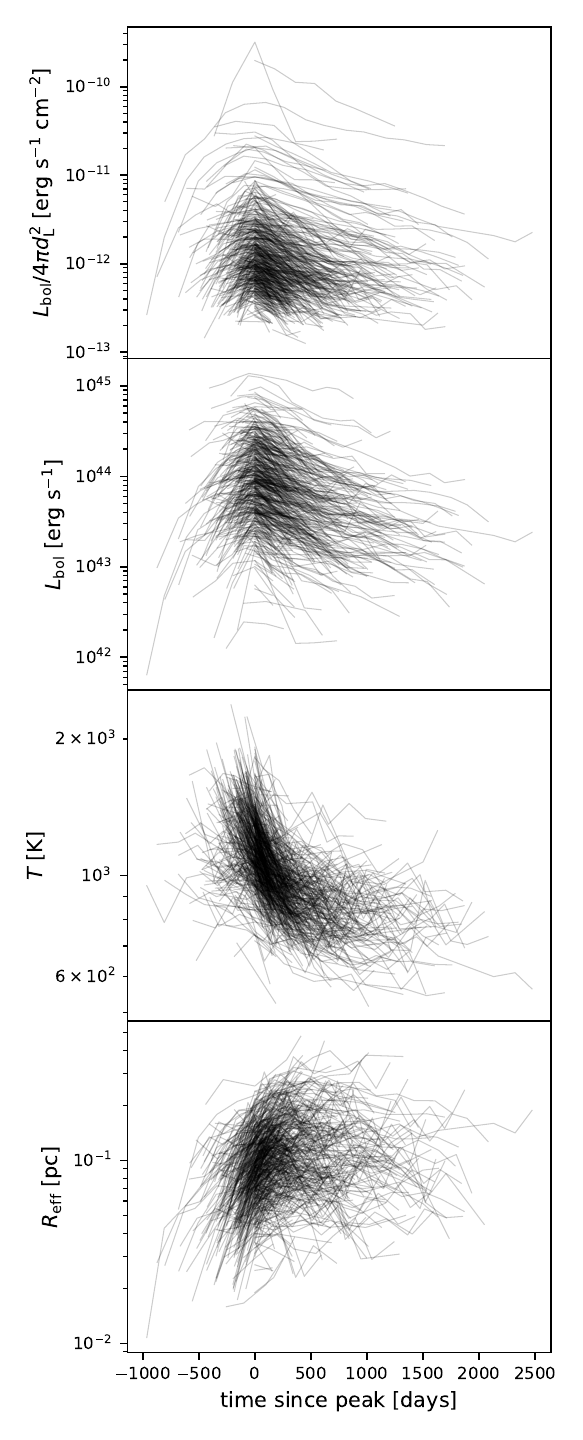}
        \caption{Parameters from the blackbody fits for 449 sources where at least two epochs have a good fit. The time is given in the source frame. \textit{First panel}: Bolometric flux based on the bolometric luminosity and the luminosity distance. \textit{Second panel}: Bolometric luminosity (see Equation \ref{eq:Lbol}). \textit{Third and fourth panel}: Temperature and effective radius (see Equation (\ref{eq:Lbb})).}
        \label{fig:sample_lc}
    \end{figure}
\begin{figure}
        \centering
        \includegraphics[width=0.49\textwidth]{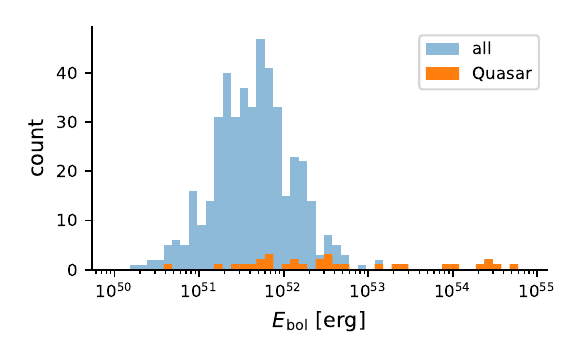}
        \caption{The distribution of the total emitted energy, assuming hot dust (Equation (\ref{eq:Ebol})). The Quasars are identified by a match to Milliquas (see Section \ref{sec:bb}).}
        \label{fig:ebol}
    \end{figure}

    We assume that the infrared flares are due to reprocessed emission by dust which is distributed in a thin, spherical shell around the central engine. We modeled the spectrum of the emission with a blackbody, although the dust emission is expected to follow a modified blackbody spectrum \citep{draineOpticalPropertiesInterstellar1984}. However, the resulting bolometric luminosity is almost unaffected while the temperature is only slightly lower \citep{jiangMidinfraredOutburstsNearby2021}. We fitted the rest frame luminosity in each band to the corresponding value predicted by the blackbody emission
    \begin{equation}
        \label{eq:Lbb}
        \nu L_{\mathrm{\nu, BB}} = \pi \cdot \nu  B_{\nu}(T) \cdot 4 \pi \, R_{\mathrm{eff}}^2
    \end{equation}
    to obtain a temperature $T$ and an effective radius $R_\mathrm{eff}$ for each visit. We numerically searched for the temperature that solves the ratio of Equation (\ref{eq:Lbb}) for the two bands. We used that to seed a maximum likelihood fit to obtain the best-fit values for $T$ and $R_\mathrm{eff}$. To investigate how well the parameter space is constrained by the data and to derive uncertainties, we performed MCMC sampling and obtained posterior distributions using \texttt{emcee} \citep{foreman-mackeyEmceeMCMCHammer2013}. Because we fit the rest frame quantity $\nu_\mathrm{e}L_{\mathrm{\nu_e}}$, we directly obtain the rest frame temperature. For each pair of fit parameters, we also calculate the bolometric luminosity
    \begin{equation}
        \label{eq:Lbol}
        L_{\mathrm{bol}} = \sigma_{\mathrm{B}} T^4 \cdot  4 \pi \, R_{\mathrm{eff}}^2.
    \end{equation}
    where $\sigma_\mathrm{B}$ is the Stefan-Boltzmann constant. To ensure a sufficient signal-to-noise ratio, we first required the luminosity ratio of the visit to be greater than its uncertainty. Secondly, we calculated the autocorrelation time $\tau$ of the MCMC ensemble \citep{sokalMonteCarloMethods1997}. We stopped the chain after $N$ iterations if $N > 100 \tau$, providing us with at least 100 independent samples per visit. If this condition is not reached after $N=10000$ iterations, the fit has not converged and we did not include the visit in the analysis. If the 69th percentile for the bolometric luminosity is larger than the best-fit value, we do not consider the sampling to give meaningful constraints. For 568 sources, we find at least two epochs with good constraints on the fit parameters.
    Figure \ref{fig:ngc7392_bb} shows the resulting blackbody spectra for the flare with the highest fluence in our sample, located in the nearby galaxy NGC 7392. The derived parameters are shown in Figure \ref{fig:ngc7392_lc}. They agree with previous measurements that suggest it as one of the nearest TDE candidates  \citep{panagiotouLuminousDustobscuredTidal2023a}.
    The evolution of $L_\mathrm{bol}$, $T$ and $R_\mathrm{eff}$ for all 568 sources with at least two epochs with good constraints is shown in Figure \ref{fig:sample_lc}, as well as the bolometric flux $L_\mathrm{bol} / 4\pi d_\mathrm{L}^2$.
    To calculate the total emitted energy, we integrate $L_\mathrm{bol}$ over the total duration of the flare in the source frame by linearly interpolating between measurements, so that
    \begin{equation}
        \label{eq:Ebol}
        E_{\mathrm{bol}} = \int L_{\mathrm{bol}}(t) \, \mathrm{d}t = \sum_{k=1}^N \frac{L_\mathrm{bol}(t_{k-1}) + L_\mathrm{bol}(t_{k})}{2} (t_k - t_{k-1})
    \end{equation}
    where the sum runs over all $N$ good fit epochs.
    The distribution is shown in Figure \ref{fig:ebol}.

\begin{figure}
        \centering
        \includegraphics[width=0.49\textwidth]{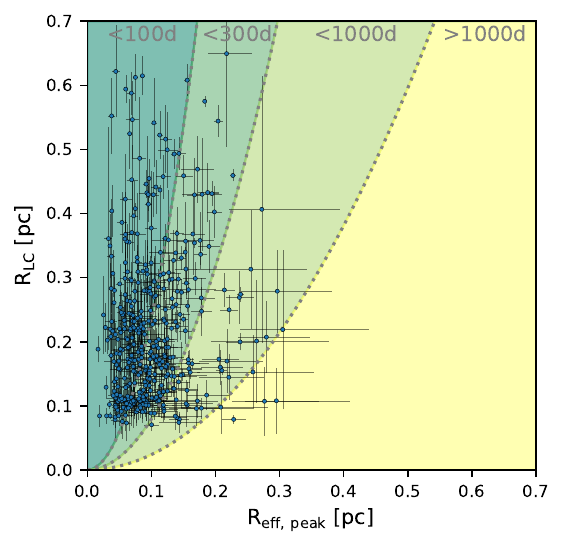}
        \caption{The dust radius estimated from the IR lighturve $R_\mathrm{LC}$ against the radius at peak $R_\mathrm{eff, \, peak}$ obtained from the blackbody fits. The colored contours represent the actual blackbody radius for a finite width of the initial O/UV transient (see Equation (\ref{eq:Rbb})).}
        \label{fig:radius}
    \end{figure}

\begin{figure*}
        \centering
        \includegraphics[width=.99\textwidth]{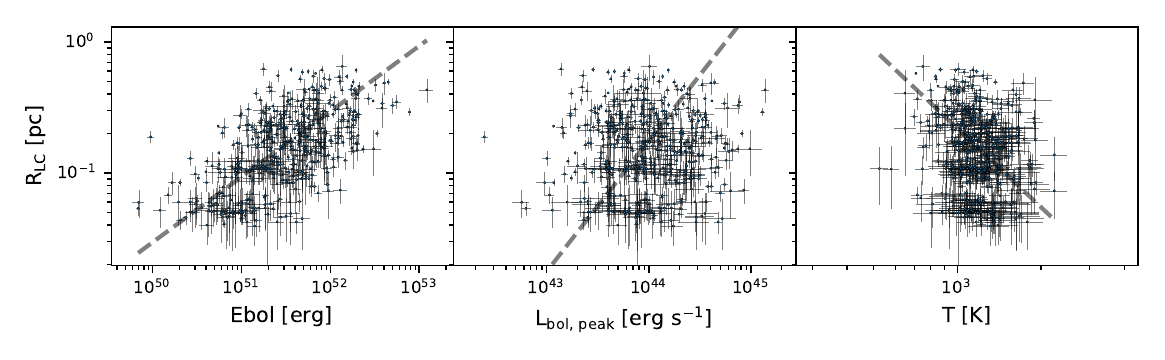}
        \caption{\textit{First Panel}: Correlation of the total emitted bolometric energy $E_\mathrm{bol}$, the peak bolometric luminosity $L_\mathrm{bol,\,peak}$ and the dust radius infered from the light curve $R_\mathrm{LC}$. The dashed lines represent the theoretically expected relations.}
        \label{fig:correlation_panel}
    \end{figure*}

    To test whether the infrared flares are dust echoes, we make sure that the derived parameters are consistent with this assumption.
    \\
    Besides hot dust, infrared emission from relativistic electrons inside jets can contribute to the variability of the MIR light curves \citep{jiangRAPIDINFRAREDVARIABILITY2012, liaoMultiwavelengthVariabilityProperties2019}. The same population of electrons typically also produces synchrotron emission in the radio band.
    To identify possible candidates for this emission, we matched our sample to the Million Quasars Catalog (Milliquas;  \citealt{fleschMillionQuasarsMilliquas2023}), the largest compilation of type-I Quasars and AGN. We found 86 matches within a search radius of \qty{5}{\arcsec} where the maximum distance of any match we found is \qty{1.4}{\arcsec}. 70 objects are marked as core-dominated Quasars, BL Lacs, or narrow-line AGN. We assume that for most of these high-redshift objects, the infrared emission comes from synchrotron emission. We exclude these from further analysis. Consequently, the blackbody fits for some of those objects give unreasonable high values for the bolometric energy $E_\mathrm{bol} \gtrsim \qty{e53}{\erg\per\second}$ (see Figure \ref{fig:ebol}).
    \\
    To validate the dust radius obtained from the blackbody fits, we can compare it to the value estimated from the light curve shape. The dust echo light curve can be obtained from the light curve of the initial transient by convolving it with a tophat function. Its width $\tau$ is determined by the radius of the dust shell $\tau = 2 \cdot R / c$, where $c$ is the speed of light \citep{velzenDISCOVERYTRANSIENTINFRARED2016}. We can relate this to the bolometric light curve by dividing the integral by the peak:
    \begin{equation}
        \tau = \frac{1}{L_\mathrm{bol,\,max}} \int L_{\mathrm{bol}}(t) \, \mathrm{d}t.
    \end{equation}
    For a tophat function, this is a good approximation. Because we defined the bolometric energy as the integral in Equation (\ref{eq:Ebol}), the radius of the dusty region determined from the light curve is then
    \begin{equation}
        \label{eq:RLC}
        R_{\mathrm{LC}} = \frac{c \, E_\mathrm{bol}}{2 \, L_\mathrm{bol,\,max}}
    \end{equation}
    If we want to compare this with the radius we get from the blackbody fits, we have to consider that at any time we only see a spatial slice of the dusty region due to the light travel time. We derived $R_\mathrm{eff}$ assuming a blackbody with an area of $A_\mathrm{BB, \, eff} = 4 \pi R_\mathrm{eff}^2$. However, because of the extension of the sphere, the surface area of the sphere that we see is $A_\mathrm{BB} \approx 2\pi R_\mathrm{BB} \cdot c\Delta T_\mathrm{opt}$ where $\Delta T_\mathrm{opt}$ is the width of the activating flare. The relation between the effective and the actual radius is
    \begin{equation}
        R_\mathrm{BB} = \frac{2}{c \Delta T_\mathrm{opt}} R_\mathrm{eff}^2.
        \label{eq:Rbb}
    \end{equation}
    The comparison between $R_\mathrm{LC}$ and the peak effective radius $R_\mathrm{eff,\, peak}$ is shown in Figure \ref{fig:radius} with an indication for different values of $\Delta T_\mathrm{opt}$ using Equation (\ref{eq:Rbb}). For typical values for the duration of TDEs \citep{komossaTidalDisruptionStars2015, gezariTidalDisruptionEvents2021} the model is self-consistent, so we can take $R_\mathrm{LC}$ as a reasonable estimate for $R_\mathrm{BB}$ \\
    Moreover, the expected relation between the dust radius and the total bolometric energy is expected to be $R_\mathrm{LC} \sim E_\mathrm{bol}^{1/2}$ \citep{velzenDISCOVERYTRANSIENTINFRARED2016}. With equation (\ref{eq:RLC}), this implies that $R_\mathrm{LC} \sim L_\mathrm{bol,peak}$. Assuming perfect blackbody emission, the relation between radius and temperature is $R_\mathrm{LC} \sim T^{-2}$. Figure \ref{fig:correlation_panel} shows these expected relations as the grey dotted line along with the measured values. There is a large scatter in all of the cases but the scaling of $E_\mathrm{bol}$ with the dust radius seems to fit the data reasonably well. Although there is no conclusion possible for the temperature and the peak bolometric luminosity, this further strengthens the dust echo model.\\
    The temperature evolution in Figure \ref{fig:sample_lc} is also consistent with cooling dust. Most of the flares start at a temperature of $T<\qty{2000}{\kelvin}$ which is broadly consistent with the sublimation temperature for the relevant dust grains \citep{velzenDISCOVERYTRANSIENTINFRARED2016}, before cooling to lower temperatures of \qtyrange{600}{800}{\kelvin}. \\
    The total emitted energy falls in the range between a few times \qtyrange{e50}{e53}{\erg} with most of the events between \qtyrange{e51}{e52}{erg} (see Figure \ref{fig:ebol}). We will discuss this in more detail in the next section, especially Section \ref{sec:tae}.

    \subsection{Possible Nature of the Initial Transient}
    \label{sec:origin}

    \subsubsection{Contribution by Active Galactic Nuclei}
    \label{sec:agn}

    Variability in active galaxies is well known as well as delayed IR emission due to dust reprocessing \citep{ulrichVARIABILITYACTIVEGALACTIC1997}. Although unusually bright IR flares compared to the extraneous part of the light curve were selected, hosts that showed AGN activity before the flare were not explicitly excluded. We can expect a certain percentage of the flares to be connected to the usual accretion onto the SMBH, so we tried to estimate their contribution to our sample.\\
    We used the classification based on Baldwin–Phillips–Terlevich (BPT) diagrams \citep{baldwinCLASSIFICATIONPARAMETERSEMISSIONLINE1981} derived from SDSS spectroscopic observation \citep{bresciaAutomatedPhysicalClassification2015} to get robust classification of the host galaxies. We find 61 matches of which 22 are star-forming galaxies, 19 composite objects, 11 AGN, and 9 LINER galaxies. So at least a third of the spectroscopically classified hosts do not show signs of AGN activity before the flare. However, this small subsample is not entirely representative as it is on average one magnitude brighter and the W1-W2-color is around \qty{.16}{\vegamag} bluer. \\
    To look for possibly classified optical counterparts, we found 68 matches on the \textit{Transient Name Server}\footnote{\url{https://www.wis-tns.org}} (TNS). While 54 are unclassified, 6 objects are classified as AGN flares. However, at least five of them are unlikely regular statistical AGN variability.
    While a regular TDE origin was rejected for AT2017bgt, it was attributed to enhanced accretion onto the SMBH, a re-ignition of the active nucleus, and was used to define a new flare class \citep{trakhtenbrotNewClassFlares2019}.
    AT2018bcb shares similarities both with regular TDEs and the AT2017bgt-like flares \citep{neustadtTDENotTDE2020}.
    This is also true for the brightest flare in our sample in the nearby AGN NGC 1566 \citep{ochmannTransientEventNGC2024}.
    The spectrum of AT2017fro was initially classified as a supernova type II \citep{brimacombeDiscoverySixASASSN2017} and later found to be more consistent with an AGN \citep{arcaviFLOYDSClassificationASASSN17jz2017}. More detailed analysis revealed similarities with other nuclear transients of uncertain nature like CSS100217 \citep{drakeDISCOVERYNATUREOPTICAL2011} and AT2019fdr \citep{frederickFamilyTreeOptical2021, reuschCandidateTidalDisruption2022}. It is found to be most compatible with a supernova type IIn in an AGN that triggered enhanced accretion activity but a TDE origin can not conclusively be ruled out \citep{holoienInvestigatingNatureLuminous2022}.
    AT2019avd is an extensively studied nuclear flare in a previously inactive galaxy that shows X-ray properties consistent with a TDE scenario although the optical and UV emission is less TDE-like \citep{malyali2019avdNovelAddition2021}. It has also been suggested as a turn-on AGN event \citep{frederickFamilyTreeOptical2021} similar to AT2017fro, as a partial TDE \citep{chen2019avdTidalDisruption2022} or a particular type of jetted TDE \citep{wangRadioDetectionAccretion2023}.
    AT2018dyk was initially classified as a TDE \citep{arcaviFLOYDSClassification2018dyk2018} but this interpretation was later disputed in favor of an AGN turn-on \citep{frederickNewClassChanginglook2019}.
    AT2022sxl was classified as an AGN flare \citep{reguittiPadovaAsiagoTransientClassification2022} but is unrelated because it happened roughly six years after the infrared peak.
    In summary, all the reported optical AGN flares in our sample are not due to standard statistical AGN variability but rather isolated events associated with the enhanced accretion onto the SMBH, while the exact nature of the trigger of this enhancement is less clear.
    \\
    To investigate the total contribution of flares happening in AGN hosts, we classified the galaxies based on their red infrared color. We do not use the measured baseline magnitudes because of the biases affecting dim sources (see Section \ref{sec:sup}). The redder band W2 is less sensitive and the bias more prominent, reddening the measured color, in particular for blue sources. So using the measured baseline leads to an overestimation of the percentage of AGN hosts. Instead, the magnitudes derived from stacked WISE images from the parent sample catalogs were used. \\
    Employing the common selection cut on the magnitude difference between the W1 and W2 bands, $m_\mathrm{W1} - m_\mathrm{W2} > 0.8$ \citep{sternMIDINFRAREDSELECTIONACTIVE2012}, the percentage of hosts with signs of AGN activity in our flare sample is 14\%. While it does increase from 6\% in the parent galaxy sample, it still shows that most of the hosts do not have strong contributions from a pre-existing AGN.
    \\


    \subsubsection{Contribution by Supernovae}
    \label{sec:sn}

    Infrared emission from hot dust after supernova explosion has been studied in the past and some supernovae remain detectable in the infrared even decades after explosion \citep{tinyanontSYSTEMATICSTUDYMIDINFRARED2016}. It is therefore worth looking for supernovae also in our sample. We found six matches with transients classified as supernovae on TNS. We suggest, however, that at least two of them could instead be flares connected to the accretion onto the SMBH.
    \\
    SN 2018gn was first classified as a type II supernova \citep{falcoSpectroscopicClassificationASASSN18am2018} but recently interpreted as a dusty TDE based on the large dust echo and the emergence of coronal lines \citep{thevenotMidinfraredDetectionsType2021, wangASASSN18apDustyTidal2024} and is part of the sample of infrared detected TDEs by the WISE Transient Pipeline \citep{mastersonNewPopulationMidinfraredselected2024}.
    SN 2020edi was classified as a superluminous supernova of type II in the center of an AGN \citep{tuckerSCATTransientClassification2021}. Its very luminous dust echo was previously reported \citep{thevenotSupernovaeDetectedNEOWISER2021} and correspondingly, our inferred bolometric energy of almost \qty{e52}{\erg} matches that of the other nuclear transients. Spectroscopic follow-up observations could determine the emergence of coronal lines, similar to SN 2018gn.
    \\
    We do not find evidence that the remaining four supernovae are instead connected to SMBH accretion. Three show spectral features interpreted as signs of interaction with the circumstellar medium and are classified as type IIn: SN 2020iq \citep{dahiwaleZTFTransientClassification2020a}, SN 2018ctj \citep{fremlingZTFTransientClassification2018} and 2018hfm \citep{zhangSN2018hfmLowenergy2022}. We note that SN 2020iq has a high inferred bolometric energy \citep{thevenotSupernovaeDetectedNEOWISER2021} of around \qty{2e51}{\erg} which is almost a magnitude higher than that of the other supernovae. Nevertheless, the distance to the closest source detected with Pan-STARRS is around \qty{0.4}{\arcsec} which puts it at an angular diameter distance of roughly \qty{700}{\parsec} from the host center, too far to be related to the SMBH.
    SN 2018ktv was classified as a type IIL \citep{fremlingZTFTransientClassification2019}.
    \\
    To also check for supernova detections before the introduction of the TNS, we crossmatched with a list of supernovae detected by ASAS-SN\footnote{\url{www.astronomy.ohio-state.edu/asassn/sn_list.html}}. Besides ASASSN-18xl/AT2018hfm, ASASSN-18ap/AT2018gn, and ASASSN-17jz/AT2017fro which are described above there were no additional matches.
    \\
    To confirm that the infrared flares are indeed happening in the center of the host galaxies, we calculated the offset of the position of the data related to the flare to the position of the baseline data. We found all three supernovae above with a separation of more than \qty{0.5}{\arcsec}. We find 36 more sources with an angular separation at least as high but for only three of them, the offset seems real. The rest likely suffer from inaccurate photometry or confusion of the host galaxies (see Appendix \ref{sec:separation}). The high angular separation suggests that these objects are supernovae, bringing the total number up to six. This implies a contamination in our sample by only around 1\%.\\
    We can conclude that the contribution of supernovae in our sample is subdominant.


    \subsubsection{Transient Accretion Events}
    \label{sec:tae}

    The previous two Sections concluded that the sample is neither dominated by supernovae nor by regular AGN activity. However, as mentioned in Section \ref{sec:sn}, the flares do happen in the center of their host galaxies. Because all derived physical parameters are consistent with the dust echo of an initial flare happening in a dust shell with $R \approx \mathcal{O}(\qty{e-1}{\parsec})$, this makes them likely dust echos of extreme accretion events onto the SMBH. The inferred bolometric energy is \qtyrange{e51}{e52}{erg} which is roughly an order of magnitude higher than the estimate for most of the (candidate) supernovae but consistent with the energy output of a TDE, assuming a covering factor of $f_c \gtrsim 0.1$, the disruption of a solar mass star, subsequent accretion of half its mass and a dissipation efficiency of $\epsilon \approx 0.1$. \\
    Indeed, roughly half of the 39 brightest flares are already classified as either TDE candidates or peculiar AGN flaring activity (see Table \ref{tab:catalog}). As mentioned in Section \ref{sec:agn}, the brightest flare in our sample in the nearby AGN NGC 1566 shares similarities with both TDEs and peculiar AGN flares \citep{ochmannTransientEventNGC2024} similar to AT2017gbt \citep{trakhtenbrotNewClassFlares2019} which is the fourth brightest flare in our sample. Both the second and third brightest flares are promising candidates for infrared-detected TDEs \citep{panagiotouLuminousDustobscuredTidal2023a, mastersonNewPopulationMidinfraredselected2024} as well as seven other flares. An additional five of the 39 brightest are part of the MIRONG sample \citep{jiangMidinfraredOutburstsNearby2021}. Also, the 14th brightest flare AT2019avd is a potential TDE \citep{malyali2019avdNovelAddition2021}. \\



    \section{Completeness}
    \label{sec:completeness}
    To make any statement about a population of astrophysical sources, it is vital to understand the fraction of sources that are missing from the underlying catalog. In principle, there are three distinct reasons for the loss: the incompleteness of survey area, depth, or limited detection efficiency in the survey volume. In the case of this study, the first point is linked to the choice of the parent galaxy sample while the selection efficiency results from the bayesian blocks analysis and subsequent cuts. The depth is influenced by both as the parent galaxy sample will be less complete at higher redshifts and also the selection efficiency is worse for fainter signals. In the following, we will focus on the detection efficiency before discussing the parent galaxy sample and depth.

    The detection efficiency is a convolution of the detection efficiency of WISE and the selection efficiency of our pipeline. Because each step would require its own set of dedicated simulations, which is beyond the scope of this paper, we instead estimate the efficiency of our selection by comparing our sample to results from the literature. The results per object are listed in Table \ref{tab:rejection_reasons}
    \\
    \citet{mastersonNewPopulationMidinfraredselected2024} identified 18 candidates for infrared-selected TDEs using the WISE Transient Pipeline (WTP). These candidates were selected based on their high peak luminosity, fast rise, and monotonic decline, with no underlying AGN host or signs of prior variability.
    Table \ref{tab:completeness} shows a summary of the results of our pipeline for this sample. We classify exactly half of these flares as dust-echo-like, while four candidates are classified as long flares. Another four get rejected because the flare is not sufficiently strong compared to the variability of the rest of the light curve. One flare does not have a baseline detection before the excess.
    Because the selection is based on a crossmatch against a galaxy catalog that extends up to about \qty{200}{\mega\parsec} \citep{cookCensusLocalUniverse2019}, the WTP TDE sample does not pick up candidates beyond a redshift of $z \lesssim 0.045$. Among the 40 brightest flares in our sample, only 13 are below this redshift. Six of those are not included in the WTP TDE sample (see Table \ref{tab:catalog}).
    \\
    The MIRONG sample \citep{jiangMidinfraredOutburstsNearby2021} followed a similar approach of collecting WISE photometry for a parent galaxy sample. The selection of candidate variable galaxies is based on the difference between the maximum and minimum brightness. Distinct outbursts are selected based on the brightness increase from the baseline to the peak. The result is a sample of 137 flares. Only 33 (24\%) of those are selected as dust-echo like by our pipeline (see Table \ref{tab:completeness}) while 40 are long flares and 35 show signs of extraneous activity (55\%). Three of the galaxies are missing from our parent galaxy sample, all of which are at a high redshift between 0.2 and 0.3 compared to the other galaxies. The remaining 26 flares (19\%) get rejected based on their data quality because we either find the excess to be due to an outlier, the peak to be during a longer gap in the data, or no baseline detection before the flare. In two cases the rejection is related to the initial selection of photometry datapoints described in Section \ref{sec:timewise}: The coordinates of SDSSJ010320+140149 are too far away from the WISE measurements, which are centered on its nucleus, so our data selection rejects the photometry as unrelated. The data points that constitute the flare in SDSSJ153151+372445 are significantly offset from the nucleus and also rejected by our photometry selection. The resulting light curve does not pass the initial variability cut on $\redchi < 1$.
    Using the parent galaxies of the MIRONG sample (SDSS spectroscopic galaxies below a redshift of $z < 0.35$ ), we verified that only four of our objects are not included in that sample. Among them are AT2018dyk and SN2018ctj which are discussed above. The two other flares in WISE J170944.87+445042.2 and WISEA J085808.47+082116.1 are faint ($\nu_\mathrm{W1}F_{\nu_\mathrm{W1},\,\mathrm{peak}} \approx \qty{e-13}{\erg\per\s\per\square\cm}$) but clearly detectable over at least three visits. Concerning the sky coverage, we can conclude that our sample represents a clean expansion of MIRONG.
    \\
    Both the previously mentioned samples and the analysis in this work rely on cataloged host galaxies. Because galaxies have to be bright enough to enable observations that are sufficient for such a classification, both intrinsically dim and far away hosts are likely missing from such catalogs. A possibility to decouple the infrared flare search from the host parameters is to rely on a different tracer of possible locations of the flare. \citet{vanvelzenEstablishingAccretionFlares2024} used nuclear flares detected in the optical by ZTF and searched for infrared flares at the same position using WISE data. The selection of the optical flare only relies on the detection of the host to determine the angular offset but not a classification. The result is a sample of 63 optical transients with large dust echos that are likely TDEs or high-amplitude AGN flares. To highlight this unification, they are dubbed accretion flares (ZTF AFs from now on). 22 of these flares are classified by our pipeline as dust-echo-like (35\%, see Table \ref{tab:completeness}). Most of the rejected events (around 27\%) do not pass the cut on extraneous activity while 14\% are classified as long flares. For AT2019pev we only find one datapoint in one band in the single exposure photometry because most of the datapoints are excluded due to de-blending (see Section \ref{sec:data}). For AT2018jut and AT2019dll we do not find an excess in both bands and for AT2020aezy no excess at all. This is because the significance of the excess is not high enough to be recognized by the bayesian blocks but high enough to pass the significance threshold when using the expected flare start time based on the optical flare. This highlights the advantage due to prior knowledge about the infrared excess time. For AT2021aeuh and AT2019msq the flux in the AllWISE single epoch photometry is systematically lower than in the NEOWISE-R photometry which leads to a false identification of the whole NEOWISE-R period as a flare. The objects are then rejected because the peak of that supposed flare would fall in the gap between 2011 and 2014.
    The hosts of another 14\% are not included in our parent galaxy sample. Those galaxies are on average one magnitude dimmer in the WISE bands and also redder, indicating that they are further away although redshift measurements are necessary for confirmation.
    \\
    The results for these samples combined suggest completeness of the selection procedure of roughly $25-50\%$, where the main losses are attributed to our definition of long flares and extraneous activity. However, we remind the reader that these cuts are necessary to exclude AGN variability as the source of the IR flares as described in Section \ref{sec:agn}.

    As previously mentioned, the completeness in absolute numbers per volume also depends on the completeness of the parent sample since it provides the positional information. Assessing its completeness is complicated, especially because it is not derived from a homogenous survey sample but from three different catalogs (see Section \ref{sec:parent_sample}). If these catalogs were homogenous over the whole sky, one could in principle obtain the completeness by comparing to a small and deep observation and extrapolate that to the rest of the sky. However, because two of the three parent sub-samples make use of Pan-STARRS data, the sky below a declination of around \ang{-30} is sampled significantly worse. Another possibility is to compare the cumulative luminosity included in our sample to a measured luminosity function. However, there exists no such measurement based on the WISE filters\footnote{Luminosity functions were derived from observations of the Spitzer Space Telescope but only for AGN \citep{lacySpitzerMidinfraredAGN2015}}. We do therefore not attempt to calculate the completeness ourselves. But because we included all NED-VLS sources, a lower estimate of our sample's completeness is given by the completeness of the NED-LVS, which is highly complete in the very nearby universe up to \qty{30}{\mega\parsec}, around 70\% complete to \qty{200}{\mega\parsec} and drops to about  1\% at \qty{1}{\giga\parsec} \citep{cookCompletenessNASAIPAC2023}. Because we include many more faint sources, we can expect that our parent sample improves on this, especially at high distances.

    \begin{table*}
        \centering
        \begin{tabular}{lrrrrrr}
\toprule
Sample & \multicolumn{2}{r}{MIRONG} & \multicolumn{2}{r}{WTP TDEs} & \multicolumn{2}{r}{ZTF AFs} \\
 & \# & [\%] & \# & [\%] & \# & [\%] \\
Result &  &  &  &  &  &  \\
\midrule
\textbf{dust echo like} & 33 & \textbf{24} & 9 & \textbf{50} & 22 & \textbf{35} \\
long flare & 40 & 29 & 4 & 22 & 9 & 14 \\
\ref{sup:strength}: extraneous activity & 35 & 26 & 4 & 22 & 17 & 27 \\
\ref{sup:coincidence}: no coincident excess region & - & - & - & - & 2 & 3 \\
\ref{sup:excess}: no excess & - & - & - & - & 1 & 2 \\
\ref{sup:baseline}: no prior baseline & 2 & 1 & 1 & 6 & - & - \\
\ref{sup:outlier}: outlier in the lightcurve & 2 & 1 & - & - & - & - \\
\ref{sup:gap}: gap in the lightcurve & 20 & 15 & - & - & 2 & 3 \\
no data & 1 & <1 & - & - & - & - \\
not in parent sample & 3 & 2 & - & - & 9 & 14 \\
$\chi^2_\mathrm{red} < 1$ & 1 & <1 & - & - & 1 & 2 \\
\bottomrule
\end{tabular}

        \caption{Absolute and relative numbers of the results of our pipeline for three reference samples: MIRONG \citep{jiangMidinfraredOutburstsNearby2021}, TDE candidates found by the WTP \citep{mastersonNewPopulationMidinfraredselected2024} and accretion flares from ZTF (ZTF AFs; \citealt{vanvelzenEstablishingAccretionFlares2024}). The first column describes which category our analysis assigned the flares. The numbers refer to the selection steps described in Section \ref{sec:sup}.}
        \label{tab:completeness}
    \end{table*}


    \section{Implications for the rate and its evolution}
    \label{sec:rate}

\begin{figure}
        \centering
        \includegraphics[width=.49\textwidth]{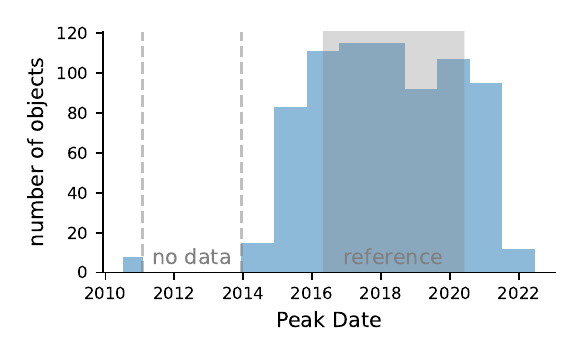}
        \caption{The distribution of peak dates over time. The reference time window from around May 2016 to June 2020 is indicated as the grey-shaded region. The break of WISE data taking is indicated with grey dashed lines.}
        \label{fig:peak_dates}
    \end{figure}

\begin{figure}
        \centering
        \includegraphics[width=.49\textwidth]{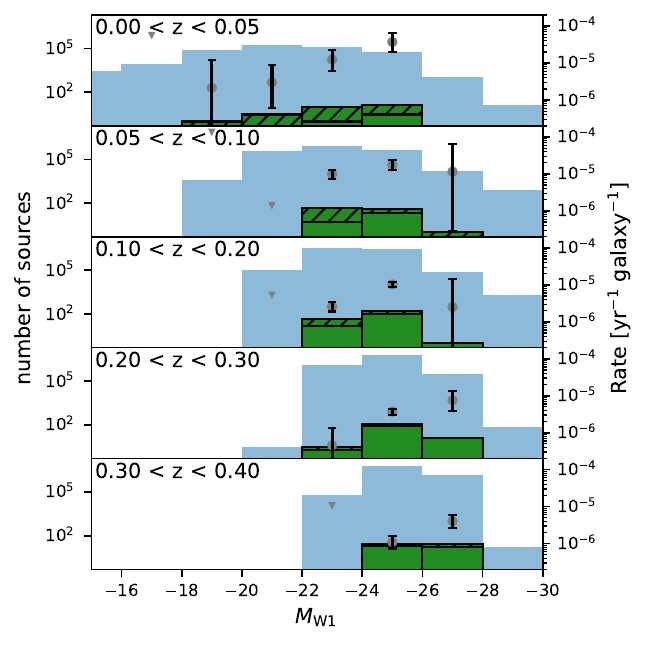}
        \caption{The rate as a function of the absolute magnitude in W1 and redshift. The light blue histogram shows the total number of galaxies in the parent galaxy sample and the green one is the host of our flare sample. The black hatched bars represent flares with $L_\mathrm{bol,\,peak} < \qty{6e43}{\erg}$. The grey data points and black errors show the rate as calculated with Equation (\ref{eq:rate}). The grey triangles are the 90\% upper limit on the rate where no flare was observed. The errors are statistical only.}
        \label{fig:rate}
    \end{figure}

\begin{figure}
        \centering
        \includegraphics[width=.49\textwidth]{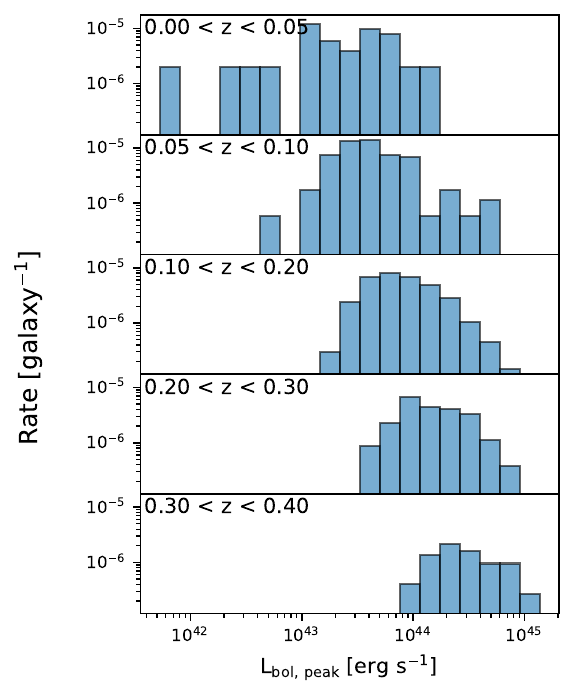}
        \caption{The distribution of the maximum bolometric luminosity as a function of redshift. }
        \label{fig:lum_fct}
    \end{figure}

    We established in Section \ref{sec:origin} that the flares in our sample are likely TDEs or at least enhanced accretion states. It is worth exploring the resulting rate of the events and comparing them with results from other samples. If the rate is in line with other TDE rate measurements it would indicate a high contribution of TDEs in our sample. Furthermore, because of the large size of our sample, we can investigate the evolution of this rate with redshift. Doing this per volume and time would require a better understanding of the completeness of our parent sample (see Section \ref{sec:completeness}) so instead we focus on the rate of events per galaxy.
    \\
    Because TDEs should only happen around black holes with masses lower than the Hills mass $m_\mathrm{BH} \approx 10^8 M_\odot$ \citep{hillsPossiblePowerSource1975}, it is useful to look at the rate as a function of the black hole mass. Because we can not directly access it, we used the host absolute magnitude in W1 $M_\mathrm{W1}$ as a proxy. There exist empirical relations between the host stellar mass and the black hole mass with different parameters for elliptical galaxies and AGN \citep{reinesRELATIONSCENTRALBLACK2015}. The host absolute magnitude in W1 $M_\mathrm{W1}$ can in turn be used as a proxy for the stellar mass \citep{kettletyGalaxyMassAssembly2018}. Combining the two relations we use $M_\mathrm{W1}$ as a proxy for $m_\mathrm{BH}$. For elliptical galaxies the Hills mass corresponds to $M_\mathrm{W1} \approx \qty{-23}{\vegamag}$ and for AGN to $M_\mathrm{W1} \approx \qty{-26}{\vegamag}$. The histograms in Figure \ref{fig:rate} show the absolute number of flares and parent galaxies as a function of redshift and $M_\mathrm{W1}$. We find that the flares happen in galaxies with $M_\mathrm{W1}$ from \qtyrange{-20.5}{-27.5}{\vegamag} with the majority between \qtyrange{-23}{-25}{\vegamag}. This would imply that most of the flares can not be due to TDEs because the SMBH is above the Hills mass using the relation for elliptical galaxies. In that case, they would be AGN flares and we would have to use the corresponding relation, resulting in a SMBH mass below the Hills mass. Because of this kind of flip-flopping, we can not conclude on the absolute magnitudes of the hosts alone. Instead, we looked at the evolution of the rate in each magnitude bin.
    \\
    The rate for magnitude bin $i$ and redshift bin $j$ is
    \begin{equation}
        \label{eq:rate}
        \mathcal{R}_{i,j} = \frac{(1 + \bar{z}_j)}{T} \frac{N_{\mathrm{flare}, i, j}}{N_{\mathrm{gal},i,j}}
    \end{equation}
    where $T$ is the sampling time, $N_{\mathrm{flare}, i, j}$ is the number of flares and $N_{\mathrm{gal},i,j}$ the number of parent galaxies in the respective bin.  $\bar{z}_j$ is the median redshift value of bin $j$ and accounts for the time dilation of the observed rate with the cosmic expansion. The uncertainty is the 95th-percentile Poisson confidence interval based on $N_{\mathrm{flare}, i, j}$. \\
    The last NEWOWISE-R data release that was included in the analysis includes data up to December 2021. Because we start with data from AllWISE, the naive estimate for the sampling time would be around $\qty{11}{\year}$ between January 2010 and December 2021. However, because the WISE satellite was inactive between February 2011 and December 2013 there is around $\qty{1.5}{\year}$ gap in the data. Also, we required a baseline detection before the flare and at least two detections of the flare (see steps \ref{sup:baseline} and \ref{sup:just_started} in Section \ref{sec:sup}). Because there is on average a visit every six months, each of these two requirements deducts about $\qty{0.5}{\year}$. The naive estimate of the sampling time would then be
    about $\qty{8}{\year}$. In reality, the effective sampling time is shorter, due to edge effects near the start, end, and break of the dataset. To include this effect, we used the average rate of flares per time $\mathcal{R}_\mathrm{ref}$ within a reference window and compared it to the overall observed rate $\mathcal{R}_\mathrm{obs}$. Assuming that the true astrophysical rate of flares is approximately constant over time, the effective sampling time is then just the total time from the start of WISE observations until the last included datapoint, scaled according to the ratio of observed to expected rate. Because the rates are just the number of observed flares $N$ over the corresponding time window, this simplifies to
    \begin{equation}
        T_{\mathrm{eff}} = T_{\mathrm{ref}} \frac{N_\mathrm{obs}}{N_\mathrm{ref}}
    \end{equation}
    Figure \ref{fig:peak_dates} shows the distribution of flare peak dates. We observe $N_\mathrm{ref}=507$ flares in our reference region from around May 2016 to June 2020 and $N_\mathrm{obs}=753$ flares in total, giving an effective sampling time of $T_\mathrm{eff} = \qty{6.1(0.3)}{\year}$. We adopt this as our sampling time $T$.\\
    The rates are shown in Figure \ref{fig:rate}. We measure the highest rate at $M_\mathrm{W1} = [-26, -24]\, \vegamag$ and $z < 0.05$ to be $\mathcal{R}_{[-26, -24], <0.05} \approx 3.7_{-1.7}^{+2.6}\times \qty{e-5}{\per\gal\per\year}$. The rate shows a rapid decrease towards brighter galaxies and a shallower towards the fainter end.
    To obtain the evolution of the rate with redshift, one needs to consider the loss of sensitivity. Figure \ref{fig:lum_fct} shows the distribution of the bolometric luminosity at peak as a function of redshift relative to the number of galaxies in the respective redshift bin. The evolution of the luminosity distribution shows the expected loss of sensitivity at low luminosities because dimmer flares are harder to detect at higher redshifts. Because we still detect flares with $L_\mathrm{bol,\,lim} \approx \qty{6e43}{\erg\per\s}$ in all redshift bins, we make the simplifying assumption, that our search is efficient in detecting flares with $L_\mathrm{bol,\,lim}$ up to the highest considered redshift bin.

\begin{figure}
        \centering
        \includegraphics[width=.49\textwidth]{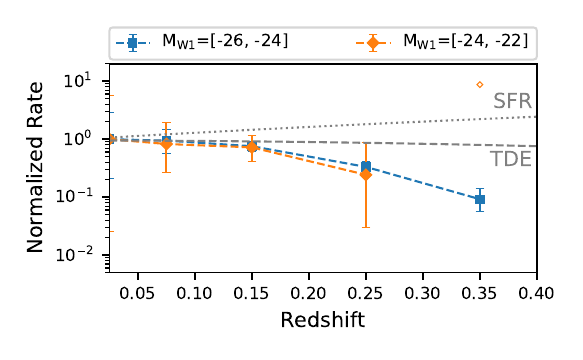}
        \caption{Evolution of the rate with redshift for the two $M_\mathrm{W1}$ bins where there are flares in the first and at least one other redshift bin, normalized to the first redshift bin. Open symbols represent upper limits. The redshift evolution of the star formation rate (SFR) \citet{madauCosmicStarFormation2014} and for TDEs \citet{sunEXTRAGALACTICHIGHENERGYTRANSIENTS2015} is shown for comparison.}
        \label{fig:redshift_rate}
    \end{figure}

    We then calculate the relative rate in redshift bin $j$
    \begin{equation}
        \mathcal{R}_j =  \frac{(1 + \bar{z}_j)}{T} \frac{\sum_{i \in \mathbb{M}} N_{\mathrm{lim}, i, j}}{\sum_{i \in \mathbb{M}} N_{\mathrm{gal},i,j}} \mathcal{R}_0^{-1},
    \end{equation}
    where $N_{\mathrm{lim}, i, j}$ is the number of flares with $L_\mathrm{bol} > L_\mathrm{bol,\,lim}$ in the respective magnitude and redshift bin. Because we are only interested in the relative evolution and not the absolute value, we only sum over the three bins $\mathbb{M}$, that contain at least one flare with $L_\mathrm{bol} > L_\mathrm{bol,\,lim}$ in any redshift bin. The relative rate is shown in Figure \ref{fig:redshift_rate} along with the relative rate for the two magnitude bins, where we found a flare in the first and at least one other redshift bin. The results for the two lower magnitude bins ($M_\mathrm{W1} = [-26, -24]\, \vegamag$ and $M_\mathrm{W1} = [-24, -22]\, \vegamag$) suggest a strictly negative evolution with a decreasing rate with higher redshifts. The evolution of the rate of TDEs based purely on the density of SMBHs predicts a shallower decline \citep{sunEXTRAGALACTICHIGHENERGYTRANSIENTS2015, kochanekTidalDisruptionEvent2016}. If TDEs do prefer post-starburst host galaxies \citep{arcaviCONTINUUMHeRICHTIDAL2014, frenchTIDALDISRUPTIONEVENTS2016, hammersteinTidalDisruptionEvent2021} which are more abundant at higher redshifts \citep{wildEvolutionPoststarburstGalaxies2016}, the expected rate could even increase although the exact evolution is unclear. As mentioned above, even for AGN the Hills mass corresponds to a host absolute magnitude of $M_\mathrm{W1} \approx \qty{-26}{\vegamag}$ which would make the flares in the highest host magnitude bin $M_\mathrm{W1} = [-26, -28] \vegamag$ more likely AGN activity than TDEs. So it would be interesting to see if the evolution of that bin is similar to the lower magnitude ones. Unfortunately, no flares above  $L_\mathrm{bol,\,lim}$ are observed in the first two redshift bins.
    \\
    Summarising, the rate seems to agree with previous results from infrared-detected TDEs and the evolution, measured for the first time, is consistent with being strictly negative. We conclude that this makes the majority of flares in our sample likely TDEs.

    \section{Discussion}
    \label{sec:discussion}

    We presented a pipeline capable of building infrared light curves based on WISE data for millions of objects and efficiently filtering them to get candidate dust echo flares. We used the pipeline together with a parent galaxy sample that covers more than three-quarters of the sky to obtain the largest sample of infrared flares compiled to date. We at least double the sky area included in the analysis compared to a previous study (MIRONG sample;  \citealt{jiangMidinfraredOutburstsNearby2021}) by not relying on SDSS observed parent galaxies, and we greatly expand the observed volume compared to the latest sample of infrared selected TDE candidates by \citet{mastersonNewPopulationMidinfraredselected2024}. The result is a sample of around 800 flares that last for thousands of days with a peak luminosity of \qtyrange{e43}{e45}{\erg\per\second} and a peak dust temperature of \qtyrange{1000}{2000}{\kelvin}. We deduced that the incident transient has a width of \qtyrange{100}{1000}{\day}. This
    is consistent with dust echoes and the inferred parameters are broadly consistent with TDEs although we can not exclude extreme accretion events in AGN. We determined the rate per galaxy which is consistent with measurements of infrared detected TDEs in the local universe \citep{mastersonNewPopulationMidinfraredselected2024}.
    The sample is deep enough to allow a preliminary study of the evolution of the rate with redshift which is consistent with a negative evolution.
    
    There are two caveats to the compilation of our sample. The biggest one remains the contribution of AGN activity. Because our selection takes into account the pre-flare variability, regular AGN variability is unlikely to be a significant contribution (see Section \ref{sec:agn}). However, although we selected dust echoes by obscured TDEs or extreme accretion events, we can not distinguish between these two scenarios based on the infrared data. An indication would be the mass of the black hole which is best accessible through spectroscopy. Observations of the host galaxies could improve our crude estimate in Section \ref{sec:rate} and determine if indeed there is a population of flares in hosts with $m_\mathrm{BH} > 10^8 M_\odot$. At least for those a TDE origin would be very unlikely. \\
    Another caveat is the Eddington bias affecting measurements near the detection limit of the single-exposure photometry (see Section \ref{sec:sup} and \ref{sec:agn}). Although we verified that the effect on the baseline measurements does not impact our result, the exact values for the luminosities of dim sources should still be taken with care. We can not exclude the possibility that the flare luminosities at higher redshifts are systematically too bright. In that case, the detection threshold would be smaller and as a result, we would include more flares at lower redshift in the evolution calculation, making the rate fall even steeper towards higher redshifts. To overcome this bias, photometry based on stacked images is necessary to make sure that also epochs are included where the source brightness is below the single-epoch limit. \\
    Finally, we note that the results for the rate and its evolution presented in this analysis do not constitute a conclusive measurement but are rather evaluating the consistency with a TDE origin. The selection efficiency especially suffers from the strict cuts necessary to distinguish regular AGN variability (see step \ref{sup:strength} in Section \ref{sec:sup}). These cuts are sensitive to the uncertainty of the baseline measurement $\sigma_\mathrm{F}$ which is small for bright flares and the resulting selection threshold is especially constraining. Consequently, the selection efficiency is likely worse for bright, nearby flares. Defining selection cuts without relying on the uncertainty of the baseline measurement could mitigate this but a detailed study is necessary to tune these new cuts. We investigated relaxing the cut on extraneous variability by requiring the dust echo strength to be greater than the significance of extraneous variability or 5, effectively capping the significance of extraneous variability. We find 329 additional flares, many of which seem to be due to non-thermal emission by quasars but are not listed in Milliquas. The brightest dust-echo-like flare that passes due to this relaxed cut is AT2019aalc, a nuclear transient claimed to be coincident with a high-energy neutrino \citep{vanvelzenEstablishingAccretionFlares2024}. We also recover AT2017gbl, a candidate obscured TDE \citep{kool2017gblDustObscured2020}. The resulting rate and evolution did not change significantly. So while the sample would be more complete and the resulting rate and evolution more robust, our conclusion would not significantly change.
    On the other hand, flares at higher redshift will appear stretched due to time dilation and might be harder to select against brightness changes on longer timescales. Correctly taking this bias into account can lead to a shallower decline towards higher redshifts.
    Furthermore, the selection effects induced by our pipeline presented in Section \ref{sec:sup} are only characterized relative to two reference samples. To evaluate the completeness of the parent galaxy sample especially as a function of redshift either an independent measurement of the galaxy luminosity functions in the WISE bands is necessary or photometric observations in bands where luminosity functions have been determined. Additionally, as previously mentioned, an absolute calibration would entail detailed simulations which is beyond the scope of this analysis.

    In summary, we demonstrated that it is possible to compile a large and deep sample of infrared flares using existing instruments. We showed that this sample is deep enough to qualitatively study the rate evolution with redshift of TDEs although more reliable photometry is necessary to obtain a quantitative result.

    \section{Reproducibility}

    The code to reproduce the analysis can be found at \url{https://gitlab.desy.de/jannisnecker/air_flares}


    \begin{acknowledgements}
        JN, TP and PMV acknowledge support by the Helmholtz Weizmann Research School on Multimessenger Astronomy, funded through the Initiative and Networking Fund of the Helmholtz Association, DESY, the Weizmann Institute, the Humboldt University of Berlin, and the University of Potsdam. AF and PMV acknowledge support from the German Science Foundation DFG, via the Collaborative Research Center SFB1491: Cosmic Interacting Matters - from Source to Signal. This research was supported by a Grant from the German-Israeli Foundation for Scientific Research and Development (GIF, Grant number I-1507-303.7/2019). This publication makes use of data products from the Near-Earth Object Wide-field Infrared Survey Explorer (NEOWISE), which is a joint project of the Jet Propulsion Laboratory/California Institute of Technology and the University of Arizona. NEOWISE is funded by the National Aeronautics and Space Administration.

        \textit{Software}: \texttt{astropy} \citep{theastropycollaborationAstropyProjectSustaining2022}, \texttt{emcee} \citep{foreman-mackeyEmceeMCMCHammer2013}, \texttt{scipy} \citep{virtanenSciPyFundamentalAlgorithms2020}, \texttt{scikit-learn} \citep{pedregosaScikitlearnMachineLearning2011}, \texttt{pandas} \citep{mckinneyDataStructuresStatistical2010}, \texttt{numpy} \citep{harrisArrayProgrammingNumPy2020}, \texttt{matplotlib} \citep{hunterMatplotlib2DGraphics2007}
    \end{acknowledgements}

    \bibliographystyle{aa}
    \bibliography{ir_flare_paper}



    \begin{appendix}


        \section{Catalogue}
        Table \ref{tab:catalog} lists the first 39 brightest flares in our sample. Below we describe each column of the full catalog which is available online.
        \begin{description}
  \item[\textbf{Name}] \hfill \\ Host Name
  \item[\textbf{RAdeg}] [$\unit{\degree}$]\hfill \\ Right Ascension in degrees
  \item[\textbf{DEdeg}] [$\unit{\degree}$]\hfill \\ Declination in degrees
  \item[\textbf{HostW1mag}] \hfill \\ Host W1 magnitude (Vega)
  \item[\textbf{HostW2mag}] \hfill \\ Host W2 magnitude (Vega)
  \item[\textbf{NEWS}] \hfill \\ Index in the NEWS sample
  \item[\textbf{AllWISE}] \hfill \\ AllWISE Designation of the Host
  \item[\textbf{PS1}] \hfill \\ Object identifier of the host in Pan-STARRS
  \item[\textbf{WISEAllSkycntr}] \hfill \\ Counter of the host in WISE AllSky
  \item[\textbf{WISEAllSkyDesignation}] \hfill \\ Designation of the host in WISE AllSky
  \item[\textbf{NEDLVSIndex}] \hfill \\ Index of the host in NED-LVS
  \item[\textbf{NEDLVSName}] \hfill \\ Name of the host in NED-LVS
  \item[\textbf{ParentSample}] \hfill \\ Index in the Flaires parent galaxy sample
  \item[\textbf{z}] \hfill \\ Redshift
  \item[\textbf{e\_z}] \hfill \\ Uncertainty of the redshift
  \item[\textbf{zsource}] \hfill \\ The source where the redshift value was taken from
  \item[\textbf{SDSSdist}] [$\unit{\arcsecond}$]\hfill \\ Distance to the associated SDSS object
  \item[\textbf{SDSSclass}] \hfill \\ BPT class of the associated SDSS object
  \item[\textbf{TNSdist}] [$\unit{\arcsecond}$]\hfill \\ Distance to the associated TNS object
  \item[\textbf{TNSobjtype}] \hfill \\ Type of the associated TNS object
  \item[\textbf{TNSname}] \hfill \\ Name of the associated TNS object
  \item[\textbf{TNSdate}] \hfill \\ Discovery date [MJD] of the associated TNS object
  \item[\textbf{milliquasdist}] [$\unit{\arcsecond}$]\hfill \\ Distance to associated Milliquas object
  \item[\textbf{milliquastype}] \hfill \\ Broad Type of the associated Milliquas object
  \item[\textbf{MirongName}] \hfill \\ Name of the host in the MIRONG sample
  \item[\textbf{WTPName}] \hfill \\ Name of the object in the WISE Transient Pipeline TDE sample
  \item[\textbf{RefTime}] [$\unit{\mjd}$]\hfill \\ Observer frame mean peak time in MJD
  \item[\textbf{x2W1}] \hfill \\ $\chi^2$ value w.r.t. the median spectral flux density in W1
  \item[\textbf{npointsW1}] \hfill \\ Number of Detections in W1
  \item[\textbf{FmedW1}] [$\unit{\milli\jansky}$]\hfill \\ Median spectral flux density in W1
  \item[\textbf{x2W2}] \hfill \\ $\chi^2$ value w.r.t. the median spectral flux density in W2
  \item[\textbf{npointsW2}] \hfill \\ Number of Detections in W2
  \item[\textbf{FmedW2}] [$\unit{\milli\jansky}$]\hfill \\ Median spectral flux density in W2
  \item[\textbf{FbslW1}] [$\unit{\milli\jansky}$]\hfill \\ Measured baseline spectral flux density in W1
  \item[\textbf{e\_FbslW1}] [$\unit{\milli\jansky}$]\hfill \\ $1\sigma$ uncertainty on FbslW1
  \item[\textbf{FbslW2}] [$\unit{\milli\jansky}$]\hfill \\ Measured baseline spectral flux density in W2
  \item[\textbf{e\_FbslW2}] [$\unit{\milli\jansky}$]\hfill \\ $1\sigma$ uncertainty on FbslW2
  \item[\textbf{startW1}] [$\unit{\mjd}$]\hfill \\ MJD of the first epoch of the determined excess in W1
  \item[\textbf{endW1}] [$\unit{\mjd}$]\hfill \\ MJD of the last epoch of the determined excess in W1
  \item[\textbf{endedW1}] \hfill \\ True if the last epoch of the excess in W1 is the last available datapoint
  \item[\textbf{startW2}] [$\unit{\mjd}$]\hfill \\ MJD of the first epoch of the determined excess in W2
  \item[\textbf{endW2}] [$\unit{\mjd}$]\hfill \\ MJD of the last epoch of the determined excess in W2
  \item[\textbf{endedW2}] \hfill \\ True if the last epoch of the excess in W2 is the last available datapoint
  \item[\textbf{strengthW1}] \hfill \\ Dust Echo strength in W1
  \item[\textbf{strengthW2}] \hfill \\ Dust Echo strength in W2
  \item[\textbf{varW1}] \hfill \\ Significance of the extraneous variability in W1
  \item[\textbf{varW2}] \hfill \\ Significance of the extraneous variability in W2
  \item[\textbf{MaxFluxW1}] [$\unit{\milli\watt\per\meter\squared}$]\hfill \\ Maxium flux density in W1
  \item[\textbf{e\_MaxFluxW1}] [$\unit{\milli\watt\per\meter\squared}$]\hfill \\ $1\sigma$ uncertainty on MaxFluxW1
  \item[\textbf{MaxFluxW2}] [$\unit{\milli\watt\per\meter\squared}$]\hfill \\ Maxium flux density in W2
  \item[\textbf{e\_MaxFluxW2}] [$\unit{\milli\watt\per\meter\squared}$]\hfill \\ $1\sigma$ uncertainty on MaxFluxW2
  \item[\textbf{FluenceW1}] [$\unit{\milli\joule\per\meter\squared}$]\hfill \\ Integrated flux density in W1
  \item[\textbf{e\_FluenceW1}] [$\unit{\milli\joule\per\meter\squared}$]\hfill \\ Uncertainty on FluenceW1 estimated by performing the integration with +/- e\_MaxFluxW1
  \item[\textbf{FluenceW2}] [$\unit{\milli\joule\per\meter\squared}$]\hfill \\ Integrated flux density in W1
  \item[\textbf{e\_FluenceW2}] [$\unit{\milli\joule\per\meter\squared}$]\hfill \\ Uncertainty on FluenceW2 estimated by performing the integration with +/- e\_MaxFluxW2
  \item[\textbf{separation}] [$\unit{\arcsecond}$]\hfill \\ Separation of flare data from baseline data
  \item[\textbf{PeakLbol}] [$\qty{e-7}{\joule\per\second}$]\hfill \\ Peak of the bolometric luminosity
  \item[\textbf{PeakTime}] [$\unit{\day}$]\hfill \\ Rest frame time of the peak of the bolometric luminosity relative to RefTime
  \item[\textbf{Ebol}] [$\qty{e-7}{\joule}$]\hfill \\ Total emitted bolometric energy
  \item[\textbf{Fluencebol}] [$\unit{\milli\joule\per\meter\squared}$]\hfill \\ Total bolometric fluence as Ebol / $4 \pi  d_L^2$
\end{description}

        \longtab[3]{
            \begin{landscape}
            \centering
            \begin{longtable}{lrrrrrrrl}
\caption{The 39 brightest mid-IR flares in the catalog, sorted by the maximum flux density in W1 $\max(\nu_\mathrm{W1}F_{\nu_\mathrm{W1}})$.The last column contains references that discuss the respective object.  All other columns are described above. (1) \cite{ochmannTransientEventNGC2024} (2) \cite{panagiotouLuminousDustobscuredTidal2023a} (3) \cite{mastersonNewPopulationMidinfraredselected2024} (4) \cite{trakhtenbrotNewClassFlares2019} (5) \cite{jiangMidinfraredOutburstsNearby2021} (6) \cite{malyali2019avdNovelAddition2021}} \label{tab:catalog} \\
\toprule
 & RA & Dec & Peak Time & $\max(\nu_\mathrm{W1}F_{\nu_\mathrm{W1}})$ & $L_\mathrm{bol, \, peak}$ & $E_\mathrm{bol}$ & $z$ & Ref. \\
 & [$\deg$] & [$\deg$] & [MJD] & [$\SI{e-12}{\erg\s^{-1}\cm^{-2}}$] & [$\SI{e43}{\erg\per\s}$] & [$\SI{e51}{\erg}$] &  &  \\
Host &  &  &  &  &  &  &  &  \\
\midrule
\endfirsthead
\caption[]{The 39 brightest mid-IR flares in the catalog, sorted by the maximum flux density in W1 $\max(\nu_\mathrm{W1}F_{\nu_\mathrm{W1}})$.The last column contains references that discuss the respective object.  All other columns are described above. (1) \cite{ochmannTransientEventNGC2024} (2) \cite{panagiotouLuminousDustobscuredTidal2023a} (3) \cite{mastersonNewPopulationMidinfraredselected2024} (4) \cite{trakhtenbrotNewClassFlares2019} (5) \cite{jiangMidinfraredOutburstsNearby2021} (6) \cite{malyali2019avdNovelAddition2021}} \\
\toprule
 & RA & Dec & Peak Time & $\max(\nu_\mathrm{W1}F_{\nu_\mathrm{W1}})$ & $L_\mathrm{bol, \, peak}$ & $E_\mathrm{bol}$ & $z$ & Ref. \\
 & [$\deg$] & [$\deg$] & [MJD] & [$\SI{e-12}{\erg\s^{-1}\cm^{-2}}$] & [$\SI{e43}{\erg\per\s}$] & [$\SI{e51}{\erg}$] &  &  \\
Host &  &  &  &  &  &  &  &  \\
\midrule
\endhead
\midrule
\multicolumn{9}{r}{Continued on next page} \\
\midrule
\endfoot
\bottomrule
\endlastfoot
NGC 1566 & 65.001750 & -54.937810 & 58324 & 117.40 & 1.90 & 0.56 & 0.00500 & 1 \\
NGC 7392 & 342.953080 & -20.608080 & 57170 & 72.42 & 5.29 & 2.78 & 0.01060 & 2, 3 \\
UGC 11487 & 297.353630 & 63.509080 & 57825 & 17.94 & 5.75 & 6.81 & 0.01900 & 3 \\
WISEA J161105.71+023400.3 & 242.773700 & 2.566750 & 58351 & 14.84 & 41.45 & 30.22 & 0.06300 & 4 \\
WISEA J010320.39+140152.5 & 15.835000 & 14.031260 & 58313 & 10.81 & 13.77 & 9.63 & 0.04230 &  \\
WISEA J034144.49-534221.4 & 55.435250 & -53.705830 & 57749 & 9.51 & 31.86 & 17.98 & 0.06660 &  \\
WISEA J130916.33-444647.2 & 197.318040 & -44.779780 & 58406 & 9.27 & 3.22 & 2.50 & 0.02490 &  \\
WISEA J154843.06+220812.6 & 237.179440 & 22.136860 & 59151 & 9.22 & 6.43 & 5.78 & 0.03130 & 5 \\
WISEA J210858.16-562831.1 & 317.242370 & -56.475330 & 58229 & 7.44 & 10.03 & 4.36 & 0.04310 & 3 \\
KUG 0143+322 & 26.676760 & 32.508030 & 58774 & 7.24 & 6.88 & 4.69 & 0.03750 & 3 \\
MCG -01-07-028 NED02 & 40.098950 & -2.728240 & 58408 & 7.00 & 4.59 & 3.30 & 0.02890 & 3 \\
WISEA J165726.80+234528.1 & 254.361710 & 23.757820 & 57979 & 6.32 & 18.40 & 17.40 & 0.05910 & 5 \\
SBS 1102+599A & 166.258250 & 59.684310 & 57964 & 6.01 & 4.60 & 5.63 & 0.03370 & 5, 3 \\
WISEA J082336.76+042302.4 & 125.903200 & 4.384020 & 59152 & 5.37 & 3.39 & 2.25 & 0.03070 & 6 \\
J023427.78-045431.0 & 38.615890 & -4.908670 & 57042 & 5.30 & - & - & - &  \\
WISEA J083812.21-584836.5 & 129.550750 & -58.810220 & 56646 & 4.32 & 8.72 & 4.18 & 0.05280 &  \\
ESO 156- G 016 & 57.220000 & -55.426360 & 57749 & 4.16 & 4.01 & 3.36 & 0.03740 & 3 \\
WISEA J003908.77-235916.4 & 9.786580 & -23.987940 & 57281 & 3.83 & 11.15 & 12.44 & 0.06440 &  \\
CGCG 047-028 & 215.725510 & 6.164820 & 57858 & 3.51 & 7.87 & 8.50 & 0.05640 & 5 \\
WISEA J074352.55-245057.8 & 115.968970 & -24.849410 & 57690 & 3.47 & 2.48 & 2.16 & 0.03240 &  \\
WISEA J115405.00-054959.9 & 178.520860 & -5.833350 & 58634 & 3.21 & 20.86 & 18.51 & 0.09040 &  \\
2MASX J14022128+3922122 & 210.588630 & 39.370110 & 57556 & 3.20 & 9.13 & 2.28 & 0.06370 & 5 \\
WISEA J154419.65-064915.3 & 236.081890 & -6.820940 & 58327 & 3.15 & 29.34 & 12.33 & 0.11110 &  \\
WISEA J034438.85+044346.2 & 56.161850 & 4.729520 & 57419 & 2.91 & 23.55 & 19.52 & 0.10060 &  \\
WISEA J104601.57-361808.4 & 161.506630 & -36.302310 & 57907 & 2.88 & 20.44 & 7.24 & 0.09700 &  \\
WISEA J113737.65-431947.3 & 174.406960 & -43.329670 & 58023 & 2.78 & 7.43 & 6.57 & 0.06140 &  \\
WISEA J082030.24-050448.5 & 125.125970 & -5.080460 & 57132 & 2.73 & 10.47 & 3.98 & 0.07340 &  \\
J205002.28+361952.6 & 312.509520 & 36.331250 & 58788 & 2.59 & - & - & - &  \\
2MASX J14024483+0726493 & 210.686620 & 7.447200 & 59595 & 2.50 & 9.50 & 4.75 & 0.07230 &  \\
WISEA J224342.87-165908.5 & 340.928650 & -16.985700 & 58265 & 2.47 & 45.24 & 12.27 & 0.14850 &  \\
WISEA J100036.44+511653.0 & 150.151860 & 51.281370 & 57960 & 2.37 & 24.89 & 13.81 & 0.11670 &  \\
J070309.40+505944.8 & 105.789210 & 50.995760 & 58552 & 2.37 & 28.38 & 16.28 & 0.12550 &  \\
WISEA J180932.24+044515.8 & 272.384170 & 4.754420 & 57194 & 2.29 & 13.67 & 15.32 & 0.08290 &  \\
WISEA J201133.97-415732.0 & 302.891580 & -41.959060 & 57306 & 2.29 & 3.82 & 0.92 & 0.04860 &  \\
J020252.75-322516.9 & 30.719800 & -32.421360 & 57385 & 2.26 & - & - & - &  \\
WISEA J222758.44+181918.5 & 336.993570 & 18.322070 & 59006 & 2.16 & 3.81 & 1.91 & 0.05060 &  \\
WISEA J104330.05+512129.0 & 160.875250 & 51.358050 & 58957 & 2.14 & 10.30 & 5.57 & 0.08160 &  \\
WISEA J215925.29+114306.7 & 329.855360 & 11.718480 & 57901 & 2.09 & 23.44 & 5.46 & 0.12100 & 5 \\
J231703.06+261401.2 & 349.262760 & 26.233710 & 58820 & 2.08 & 54.71 & 17.45 & 0.17860 &  \\
\end{longtable}

            \end{landscape}
        }


        \section{Reference Sample}

        Table \ref{tab:rejection_reasons} lists the individual flares from the samples introduced in Section \ref{tab:completeness}. The first column gives the name of the reference sample and the second column is the identifier in that sample. Columns three, four, and five are the dust echo strength, significance of extraneous variability, and $\redchi$ as determined by our analysis. The last column describes the resulting tag in our analysis, referencing the corresponding selection step in Section \ref{sec:sup} where applicable.

        \longtab[2]{
            \centering
            \begin{longtable}{llrrrl}
\caption{Reasons for rejection of reference sample objects} \label{tab:rejection_reasons} \\
\toprule
 &  & $\Delta F / F_\mathrm{rms}$ & $F_\mathrm{rms} / \sigma_\mathrm{F}$ & $\chi^2_\mathrm{red}$ & Result \\
Sample & Name &  &  &  &  \\
\midrule
\endfirsthead
\caption[]{Reasons for rejection of reference sample objects} \\
\toprule
 &  & $\Delta F / F_\mathrm{rms}$ & $F_\mathrm{rms} / \sigma_\mathrm{F}$ & $\chi^2_\mathrm{red}$ & Result \\
Sample & Name &  &  &  &  \\
\midrule
\endhead
\midrule
\multicolumn{6}{r}{Continued on next page} \\
\midrule
\endfoot
\bottomrule
\endlastfoot
\multirow[t]{137}{*}{MIRONG} & SDSSJ000046.47+143813.0 & 7.235460 & 6.176324 & 51.499871 & long flare \\
 & SDSSJ002701.04+071357.6 & 9.416179 & 5.010366 & 25.979449 & long flare \\
 & SDSSJ004500.49-004723.0 & 4.987654 & 5.604017 & 19.852376 & \ref{sup:strength}: extraneous activity \\
 & SDSSJ010320.42+140149.8 & - & - & - & no data \\
 & SDSSJ012048.00-082918.4 & 5.290421 & 12.039740 & 253.113113 & \ref{sup:strength}: extraneous activity \\
 & SDSSJ012100.68+140517.3 & 5.647730 & 7.327357 & 20.514548 & \ref{sup:strength}: extraneous activity \\
 & SDSSJ015804.75-005221.9 & 55.093009 & 5.044866 & 536.118395 & long flare \\
 & SDSSJ020552.15+000411.7 & 39.568772 & 6.698829 & 506.233210 & \textbf{dust echo like} \\
 & SDSSJ074547.87+265538.0 & 6.065738 & 6.103610 & 241.904915 & \ref{sup:strength}: extraneous activity \\
 & SDSSJ075709.70+190842.8 & 10.424613 & 3.903862 & 51.304172 & long flare \\
 & SDSSJ081121.40+405451.8 & 15.843710 & 7.293818 & 666.367277 & long flare \\
 & SDSSJ081403.78+261144.3 & 10.676061 & 5.313606 & 55.389235 & \ref{sup:gap}: gap in the lightcurve \\
 & SDSSJ081451.88+533732.6 & 34.289755 & 1.296370 & 50.196344 & \ref{sup:gap}: gap in the lightcurve \\
 & SDSSJ083536.49+493542.7 & 17.811010 & 5.238040 & 171.993866 & long flare \\
 & SDSSJ083721.86+414342.1 & 12.243468 & 7.497678 & 52.028684 & long flare \\
 & SDSSJ084157.99+052605.8 & 8.819825 & 7.603010 & 70.854667 & long flare \\
 & SDSSJ084232.88+235719.7 & 6.412621 & 9.477398 & 24.187445 & \ref{sup:strength}: extraneous activity \\
 & SDSSJ084752.78+514236.2 & 5.395527 & 5.175775 & 11.194459 & long flare \\
 & SDSSJ085434.66+111334.8 & 13.042479 & 7.318329 & 125.188190 & long flare \\
 & SDSSJ085835.91+412113.9 & 4.259047 & 5.779492 & 9.223576 & \ref{sup:strength}: extraneous activity \\
 & SDSSJ085959.47+092225.7 & 32.584427 & 3.778876 & 539.141133 & \textbf{dust echo like} \\
 & SDSSJ090924.55+192004.8 & 9.701734 & 5.078224 & 119.042151 & \textbf{dust echo like} \\
 & SDSSJ091531.04+481407.7 & 11.704594 & 4.957337 & 97.747321 & long flare \\
 & SDSSJ093135.48+662652.2 & 6.264897 & 4.142224 & 31.024734 & long flare \\
 & SDSSJ093608.59+061525.4 & 16.196292 & 2.815029 & 17.307738 & \textbf{dust echo like} \\
 & SDSSJ094303.26+595809.5 & 5.408209 & 5.323915 & 12.650494 & long flare \\
 & SDSSJ094456.57+310552.2 & 5.220873 & 9.704360 & 34.462898 & \ref{sup:strength}: extraneous activity \\
 & SDSSJ095754.77+020711.2 & 6.273216 & 4.991772 & 52.640732 & \ref{sup:gap}: gap in the lightcurve \\
 & SDSSJ100120.37+182926.7 & 4.170072 & 8.954458 & 7.178640 & \ref{sup:strength}: extraneous activity \\
 & SDSSJ100256.90+442457.8 & 9.133122 & 4.655360 & 10.924204 & long flare \\
 & SDSSJ100350.98+020227.7 & 14.319672 & 3.776481 & 78.438508 & \textbf{dust echo like} \\
 & SDSSJ100809.03+154951.4 & 19.496603 & 3.069665 & 105.320703 & \ref{sup:gap}: gap in the lightcurve \\
 & SDSSJ100931.71+343604.8 & 5.840619 & 7.921774 & 21.541772 & \ref{sup:strength}: extraneous activity \\
 & SDSSJ100955.71+220949.3 & 9.839760 & 9.975518 & 57.164248 & \ref{sup:gap}: gap in the lightcurve \\
 & SDSSJ101157.64+534857.9 & - & - & - & not in parent sample \\
 & SDSSJ101708.95+122412.2 & 3.524767 & 16.669487 & 47.759490 & \ref{sup:strength}: extraneous activity \\
 & SDSSJ102017.72+251554.4 & 7.700326 & 5.184621 & 12.675709 & \textbf{dust echo like} \\
 & SDSSJ102934.89+252635.8 & 13.488708 & 5.564740 & 28.956184 & \ref{sup:baseline}: no prior baseline \\
 & SDSSJ102959.96+482937.9 & 6.251344 & 13.197655 & 99.358152 & \ref{sup:strength}: extraneous activity \\
 & SDSSJ103753.69+391249.7 & 11.407591 & 4.859846 & 45.388561 & long flare \\
 & SDSSJ104138.80+341253.6 & 7.691065 & 5.175065 & 39.831729 & \textbf{dust echo like} \\
 & SDSSJ104306.57+271602.2 & 7.525025 & 6.519724 & 98.379477 & \textbf{dust echo like} \\
 & SDSSJ104609.62+165511.5 & - & - & - & not in parent sample \\
 & SDSSJ105145.48+210132.2 & 5.666400 & 9.385510 & 23.403331 & \ref{sup:strength}: extraneous activity \\
 & SDSSJ105344.14+552405.7 & 2.848154 & 8.681730 & 16.270909 & \ref{sup:strength}: extraneous activity \\
 & SDSSJ105801.53+544437.0 & 7.568631 & 4.630113 & 36.149122 & long flare \\
 & SDSSJ110501.98+594103.6 & 39.665225 & 6.202336 & 1511.553381 & \textbf{dust echo like} \\
 & SDSSJ110958.35+370809.7 & 44.304717 & 2.219532 & 183.171075 & \ref{sup:gap}: gap in the lightcurve \\
 & SDSSJ111122.44+592334.3 & 22.176583 & 2.429040 & 60.373333 & \textbf{dust echo like} \\
 & SDSSJ111431.84+405613.8 & 44.444486 & 1.256672 & 160.798730 & \ref{sup:gap}: gap in the lightcurve \\
 & SDSSJ111536.57+054449.7 & 18.460730 & 5.601628 & 346.176550 & \textbf{dust echo like} \\
 & SDSSJ112018.32+193345.8 & 13.902317 & 7.711235 & 171.920453 & \textbf{dust echo like} \\
 & SDSSJ112238.85+143348.4 & 4.805302 & 19.970664 & 82.419412 & \ref{sup:strength}: extraneous activity \\
 & SDSSJ112446.22+045525.4 & 24.807607 & 2.549719 & 123.793682 & long flare \\
 & SDSSJ112916.13+513123.5 & 7.890306 & 6.872648 & 34.293054 & long flare \\
 & SDSSJ113355.94+670107.1 & 18.300889 & 6.037712 & 507.653610 & \ref{sup:gap}: gap in the lightcurve \\
 & SDSSJ113901.27+613408.7 & 11.103334 & 5.650358 & 70.008886 & long flare \\
 & SDSSJ114922.03+544151.4 & 4.504857 & 5.856587 & 142.418867 & \ref{sup:strength}: extraneous activity \\
 & SDSSJ115205.33+485050.0 & 50.118754 & 2.486997 & 442.994482 & \ref{sup:gap}: gap in the lightcurve \\
 & SDSSJ115326.76+403719.2 & 16.255859 & 5.822790 & 95.590271 & \textbf{dust echo like} \\
 & SDSSJ120057.93+064823.1 & 8.561520 & 17.287579 & 246.516277 & \ref{sup:strength}: extraneous activity \\
 & SDSSJ120145.97+352522.5 & 42.640895 & 1.865613 & 173.707928 & \ref{sup:gap}: gap in the lightcurve \\
 & SDSSJ120338.30+585911.9 & 11.833001 & 2.732505 & 72.328231 & \textbf{dust echo like} \\
 & SDSSJ120842.70+330523.1 & 2.691464 & 8.594059 & 10.803171 & \ref{sup:strength}: extraneous activity \\
 & SDSSJ120942.23+320258.8 & 23.750284 & 9.475727 & 379.246307 & \ref{sup:baseline}: no prior baseline \\
 & SDSSJ121130.31+404743.2 & 6.307153 & 5.388289 & 73.887041 & \ref{sup:gap}: gap in the lightcurve \\
 & SDSSJ121457.42+101418.2 & 14.902525 & 5.708388 & 51.276918 & \ref{sup:gap}: gap in the lightcurve \\
 & SDSSJ121825.51+295154.9 & 41.000503 & 1.865143 & 278.278463 & \textbf{dust echo like} \\
 & SDSSJ121907.89+051645.7 & 15.330164 & 5.671076 & 89.094686 & \textbf{dust echo like} \\
 & SDSSJ122823.87+361729.1 & 20.547117 & 5.483538 & 142.763904 & \textbf{dust echo like} \\
 & SDSSJ123852.88+081512.1 & 6.069469 & 4.707630 & 125.426539 & \ref{sup:gap}: gap in the lightcurve \\
 & SDSSJ124255.37+253728.0 & 13.175970 & 4.156801 & 45.506969 & \textbf{dust echo like} \\
 & SDSSJ124521.42-014735.5 & 4.776322 & 5.924130 & 13.054067 & long flare \\
 & SDSSJ130355.94+220338.7 & 7.525088 & 12.523780 & 24.659995 & \ref{sup:strength}: extraneous activity \\
 & SDSSJ130532.91+395337.9 & 23.287563 & 3.012912 & 120.724890 & \textbf{dust echo like} \\
 & SDSSJ130815.58+042909.6 & 15.568442 & 4.941736 & 114.815733 & long flare \\
 & SDSSJ131022.78+251809.3 & 11.829497 & 5.084214 & 38.121376 & long flare \\
 & SDSSJ131509.34+072737.7 & 15.796398 & 8.519733 & 110.744750 & \textbf{dust echo like} \\
 & SDSSJ132259.95+330121.9 & 9.922793 & 5.485206 & 38.644442 & long flare \\
 & SDSSJ132848.46+275227.8 & 3.899487 & 20.167541 & 134.132646 & \ref{sup:strength}: extraneous activity \\
 & SDSSJ132902.05+234108.4 & 13.974117 & 5.057244 & 163.485784 & long flare \\
 & SDSSJ133212.63+203638.0 & 309.273783 & 0.626732 & 438.228317 & \textbf{dust echo like} \\
 & SDSSJ133731.36+003529.0 & - & - & - & not in parent sample \\
 & SDSSJ134032.49+184218.6 & 3.992887 & 5.683695 & 16.722340 & \ref{sup:strength}: extraneous activity \\
 & SDSSJ134105.98-004902.6 & 14.053779 & 5.164794 & 135.278992 & \ref{sup:gap}: gap in the lightcurve \\
 & SDSSJ134123.21+151650.5 & 7.636517 & 6.827178 & 56.799892 & long flare \\
 & SDSSJ134849.39+155902.1 & 26.355265 & 2.360722 & 63.378168 & long flare \\
 & SDSSJ135241.37+000925.8 & 9.652395 & 2.819655 & 20.870745 & \ref{sup:gap}: gap in the lightcurve \\
 & SDSSJ140221.27+392212.4 & 101.749204 & 4.832630 & 952.400900 & \textbf{dust echo like} \\
 & SDSSJ140648.44+062834.8 & 6.613004 & 9.104817 & 40.956218 & \ref{sup:strength}: extraneous activity \\
 & SDSSJ140950.27+105740.3 & 4.571289 & 9.851900 & 40.481812 & \ref{sup:strength}: extraneous activity \\
 & SDSSJ141235.90+411458.6 & 5.822121 & 7.487867 & 29.179248 & \ref{sup:strength}: extraneous activity \\
 & SDSSJ142254.12+060953.4 & 13.936775 & 7.827294 & 321.202300 & \textbf{dust echo like} \\
 & SDSSJ142420.78+624916.6 & 3.236530 & 38.492208 & 275.307065 & \ref{sup:strength}: extraneous activity \\
 & SDSSJ142808.89-023124.9 & 5.854497 & 6.932831 & 12.347123 & \ref{sup:strength}: extraneous activity \\
 & SDSSJ143016.06+230344.5 & 25.783246 & 4.861648 & 369.158601 & long flare \\
 & SDSSJ144024.32+175852.7 & 5.403224 & 4.179069 & 16.210109 & long flare \\
 & SDSSJ144227.59+555846.4 & 183.048216 & 3.653513 & 13421.909636 & long flare \\
 & SDSSJ144758.41+402335.9 & 15.614246 & 11.811118 & 394.460979 & \textbf{dust echo like} \\
 & SDSSJ144829.01+113732.2 & 4.316780 & 7.214739 & 78.442872 & \ref{sup:strength}: extraneous activity \\
 & SDSSJ150440.38+010735.8 & 37.343165 & 4.253255 & 1456.140895 & \ref{sup:gap}: gap in the lightcurve \\
 & SDSSJ150844.23+260249.1 & 5.758626 & 4.968898 & 30.026187 & long flare \\
 & SDSSJ151117.94+221428.2 & 7.311309 & 8.573347 & 91.304618 & \ref{sup:strength}: extraneous activity \\
 & SDSSJ151257.19+280937.5 & 5.724501 & 5.885187 & 27.818087 & long flare \\
 & SDSSJ151345.77+311125.1 & 47.166524 & 2.948985 & 244.165767 & \textbf{dust echo like} \\
 & SDSSJ152438.15+531458.6 & 19.154969 & 8.903445 & 127.762977 & \textbf{dust echo like} \\
 & SDSSJ153151.42+372445.8 & - & - & 0.719872 & $\chi^2_\mathrm{red} < 1$ \\
 & SDSSJ153310.03+272920.3 & 8.679626 & 19.731642 & 531.791620 & \ref{sup:strength}: extraneous activity \\
 & SDSSJ153711.31+581420.3 & 12.550327 & 16.313233 & 999.639717 & \ref{sup:strength}: extraneous activity \\
 & SDSSJ154029.29+005437.3 & 5.384957 & 3.442214 & 5.424775 & \ref{sup:outlier}: outlier in the lightcurve \\
 & SDSSJ154158.64+071836.4 & 8.508220 & 4.139101 & 23.421340 & long flare \\
 & SDSSJ154843.07+220812.6 & 659.673932 & 3.514871 & 12702.736785 & \textbf{dust echo like} \\
 & SDSSJ154955.20+332751.9 & 30.868220 & 5.776725 & 379.537908 & \textbf{dust echo like} \\
 & SDSSJ155437.26+525526.4 & 1271.279330 & 0.143139 & 657.535016 & long flare \\
 & SDSSJ155438.40+163637.7 & 14.749947 & 6.228978 & 29.971927 & \ref{sup:outlier}: outlier in the lightcurve \\
 & SDSSJ155440.26+362952.1 & 16.760530 & 5.629148 & 290.076472 & \ref{sup:gap}: gap in the lightcurve \\
 & SDSSJ155539.96+212005.7 & 41.948255 & 2.775905 & 258.965049 & long flare \\
 & SDSSJ155640.32+451338.4 & 18.968964 & 4.807138 & 230.319451 & \ref{sup:gap}: gap in the lightcurve \\
 & SDSSJ155743.53+272753.0 & 2.588124 & 5.552205 & 10.200410 & \ref{sup:strength}: extraneous activity \\
 & SDSSJ160052.27+461243.0 & 11.793526 & 6.265247 & 75.773865 & long flare \\
 & SDSSJ161258.17+141617.5 & 17.550276 & 5.854813 & 105.504794 & long flare \\
 & SDSSJ162034.99+240726.6 & 61.299406 & 3.576756 & 639.213949 & \textbf{dust echo like} \\
 & SDSSJ162810.05+481047.7 & 6.826520 & 3.609516 & 26.329869 & long flare \\
 & SDSSJ163246.85+441618.6 & 24.381124 & 4.926145 & 419.247326 & long flare \\
 & SDSSJ164754.38+384341.9 & 74.538679 & 3.283795 & 594.017860 & \textbf{dust echo like} \\
 & SDSSJ165726.81+234528.1 & 237.869626 & 4.502757 & 14781.705683 & \textbf{dust echo like} \\
 & SDSSJ165922.66+204947.5 & 3.924276 & 11.771531 & 34.284158 & \ref{sup:strength}: extraneous activity \\
 & SDSSJ211529.89-001107.0 & 6.509595 & 7.194816 & 11.941335 & \ref{sup:strength}: extraneous activity \\
 & SDSSJ214142.91-085702.4 & 10.083664 & 5.490241 & 66.055054 & \ref{sup:gap}: gap in the lightcurve \\
 & SDSSJ214603.88+104128.7 & 16.339630 & 5.191079 & 39.573875 & \textbf{dust echo like} \\
 & SDSSJ215055.73-010654.2 & 33.545180 & 4.658560 & 175.679014 & \textbf{dust echo like} \\
 & SDSSJ215648.47+004110.7 & 4.685745 & 10.900456 & 16.319207 & \ref{sup:strength}: extraneous activity \\
 & SDSSJ220349.24+112433.0 & 17.582776 & 5.065091 & 368.421979 & long flare \\
 & SDSSJ221541.61-010721.1 & 3.371077 & 6.499031 & 29.575881 & \ref{sup:strength}: extraneous activity \\
 & SDSSJ231055.38+222008.6 & 5.167343 & 10.500322 & 22.236243 & \ref{sup:strength}: extraneous activity \\
 & SDSSJ231222.78+133538.8 & 5.958648 & 5.588821 & 22.233304 & long flare \\
 & SDSSJ232452.26+154251.1 & 14.606632 & 5.909944 & 46.370852 & \ref{sup:gap}: gap in the lightcurve \\
\multirow[t]{18}{*}{WTP TDEs} & WTP14abnpgk & 4.971051 & 6.232620 & 91.249217 & \ref{sup:strength}: extraneous activity \\
 & WTP14acnjbu & 8.174150 & 14.032765 & 236.938983 & \ref{sup:strength}: extraneous activity \\
 & WTP14adbjsh & 59.113039 & 6.146779 & 1283.181653 & \textbf{dust echo like} \\
 & WTP14adbwvs & 12.525113 & 3.867966 & 104.866597 & long flare \\
 & WTP14adeqka & 51.673148 & 12.767377 & 5867.342130 & \textbf{dust echo like} \\
 & WTP15abymdq & 73.177181 & 5.642893 & 2594.497185 & \textbf{dust echo like} \\
 & WTP15acbgpn & 39.665225 & 6.202336 & 1511.553381 & \textbf{dust echo like} \\
 & WTP15acbuuv & 44.479987 & 5.618758 & 1666.751397 & \ref{sup:baseline}: no prior baseline \\
 & WTP16aaqrcr & 15.088180 & 6.358838 & 55.112767 & long flare \\
 & WTP16aatsnw & 3.924276 & 11.771531 & 34.284158 & \ref{sup:strength}: extraneous activity \\
 & WTP17aaldjb & 15.568442 & 4.941736 & 114.815733 & long flare \\
 & WTP17aalzpx & 10.001899 & 7.965015 & 95.018065 & long flare \\
 & WTP17aamoxe & 44.621840 & 10.767805 & 915.644330 & \textbf{dust echo like} \\
 & WTP17aamzew & 5.220873 & 9.704360 & 34.462898 & \ref{sup:strength}: extraneous activity \\
 & WTP17aanbso & 11.833001 & 2.732505 & 72.328231 & \textbf{dust echo like} \\
 & WTP18aajkmk & 83.110764 & 6.096719 & 1775.218803 & \textbf{dust echo like} \\
 & WTP18aamced & 14.931887 & 9.411465 & 167.557396 & \textbf{dust echo like} \\
 & WTP18aampwj & 73.452458 & 7.503002 & 1792.968383 & \textbf{dust echo like} \\
\multirow[t]{63}{*}{ZTF AFs} & AT2013kp & 78.258921 & 4.135601 & 1103.236541 & long flare \\
 & AT2016eix & 8.998844 & 6.760484 & 33.569221 & long flare \\
 & AT2018dyk & 14.931887 & 9.411465 & 167.557396 & \textbf{dust echo like} \\
 & AT2018ige & 12.981374 & 4.641984 & 76.570456 & \textbf{dust echo like} \\
 & AT2018iql & 30.078050 & 2.446704 & 33.652369 & \textbf{dust echo like} \\
 & AT2018jut & 1.844687 & 2.869888 & 1.779824 & \ref{sup:coincidence}: no coincident excess region \\
 & AT2018kox & 6.687261 & 8.802474 & 19.021367 & \ref{sup:strength}: extraneous activity \\
 & AT2018lcp & 5.924980 & 5.038824 & 15.961950 & long flare \\
 & AT2018lhv & 17.236311 & 3.967455 & 59.900819 & \textbf{dust echo like} \\
 & AT2018lof & 3.212749 & 9.202141 & 10.283567 & \ref{sup:strength}: extraneous activity \\
 & AT2018lzs & 3.424038 & 5.607693 & 3.797937 & \ref{sup:strength}: extraneous activity \\
 & AT2019aalc & 16.178021 & 29.512714 & 2039.271256 & \ref{sup:strength}: extraneous activity \\
 & AT2019aame & 4.145465 & 7.868080 & 7.681842 & \ref{sup:strength}: extraneous activity \\
 & AT2019aamf & 6.150682 & 8.161504 & 28.706692 & \ref{sup:strength}: extraneous activity \\
 & AT2019aamg & 7.893936 & 8.338644 & 12.922134 & \textbf{dust echo like} \\
 & AT2019aamh & - & - & - & not in parent sample \\
 & AT2019aami & 28.339675 & 4.180189 & 92.996992 & long flare \\
 & AT2019avd & 115.478889 & 3.324406 & 694.058150 & \textbf{dust echo like} \\
 & AT2019brs & 21.896789 & 6.646403 & 140.781792 & \textbf{dust echo like} \\
 & AT2019cle & 28.265672 & 4.428900 & 111.836658 & \textbf{dust echo like} \\
 & AT2019cyq & 17.336724 & 4.953578 & 42.660725 & \textbf{dust echo like} \\
 & AT2019dll & 2.699543 & 6.309565 & 1.825533 & \ref{sup:coincidence}: no coincident excess region \\
 & AT2019dqv & 69.361343 & 6.047845 & 1469.749062 & \textbf{dust echo like} \\
 & AT2019dsg & 74.945459 & 3.992322 & 449.400951 & \textbf{dust echo like} \\
 & AT2019dzh & 5.708300 & 7.620262 & 20.090393 & \ref{sup:strength}: extraneous activity \\
 & AT2019fdr & - & - & - & not in parent sample \\
 & AT2019gur & 20.219231 & 8.223698 & 78.136849 & \textbf{dust echo like} \\
 & AT2019hbh & 9.181261 & 5.758469 & 296.091001 & \textbf{dust echo like} \\
 & AT2019hdy & - & - & - & not in parent sample \\
 & AT2019idm & 36.587385 & 2.950314 & 166.409069 & \textbf{dust echo like} \\
 & AT2019ihu & - & - & - & not in parent sample \\
 & AT2019ihv & 3.485036 & 16.348000 & 22.663403 & \ref{sup:strength}: extraneous activity \\
 & AT2019kqu & 6.596225 & 9.466918 & 66.630822 & \ref{sup:strength}: extraneous activity \\
 & AT2019meh & 43.997366 & 5.994053 & 543.008547 & long flare \\
 & AT2019msq & 10.998912 & 3.649930 & 23.684240 & \ref{sup:gap}: gap in the lightcurve \\
 & AT2019mss & 17.014102 & 5.788265 & 43.704722 & \textbf{dust echo like} \\
 & AT2019nna & 8.644302 & 8.329182 & 30.209895 & \textbf{dust echo like} \\
 & AT2019nni & 6.253971 & 8.709367 & 27.787737 & \ref{sup:strength}: extraneous activity \\
 & AT2019pev & - & - & - & $\chi^2_\mathrm{red} < 1$ \\
 & AT2019qpt & 6.206294 & 7.391323 & 43.749811 & \ref{sup:strength}: extraneous activity \\
 & AT2019thh & 82.094303 & 5.167433 & 753.972348 & \textbf{dust echo like} \\
 & AT2019wrd & 15.061761 & 4.644523 & 29.172083 & long flare \\
 & AT2019xgg & 4.882445 & 6.480683 & 10.520460 & \ref{sup:strength}: extraneous activity \\
 & AT2020aezy & 1.973578 & 4.987611 & 2.407830 & \ref{sup:excess}: no excess \\
 & AT2020aezz & 8.076402 & 10.318643 & 73.891077 & \ref{sup:strength}: extraneous activity \\
 & AT2020afaa & 4.546167 & 3.065439 & 5.057972 & \textbf{dust echo like} \\
 & AT2020afab & - & - & - & not in parent sample \\
 & AT2020afac & 9.259604 & 9.365964 & 21.928723 & \ref{sup:strength}: extraneous activity \\
 & AT2020afad & - & - & - & not in parent sample \\
 & AT2020afae & 8.423922 & 10.185686 & 42.966593 & \ref{sup:strength}: extraneous activity \\
 & AT2020atq & 35.855980 & 6.917691 & 242.942820 & \textbf{dust echo like} \\
 & AT2020hle & 22.608236 & 6.985780 & 70.301757 & long flare \\
 & AT2020iq & 38.331778 & 4.271493 & 122.568764 & \textbf{dust echo like} \\
 & AT2020mw & - & - & - & not in parent sample \\
 & AT2021aetz & 33.545180 & 4.658560 & 175.679014 & \textbf{dust echo like} \\
 & AT2021aeud & 5.584794 & 4.331697 & 7.296335 & long flare \\
 & AT2021aeue & 2.398531 & 10.688374 & 6.055754 & \ref{sup:strength}: extraneous activity \\
 & AT2021aeuf & 7.269850 & 5.060153 & 32.370354 & \textbf{dust echo like} \\
 & AT2021aeug & 4.039072 & 13.195234 & 68.342928 & \ref{sup:strength}: extraneous activity \\
 & AT2021aeuh & 16.924663 & 3.375365 & 89.581822 & \ref{sup:gap}: gap in the lightcurve \\
 & AT2021aeui & - & - & - & not in parent sample \\
 & AT2021aeuj & 9.010752 & 3.961792 & 47.369657 & long flare \\
 & AT2021aeuk & - & - & - & not in parent sample \\
\end{longtable}

        }


        \section{Flares with high angular separation}
        \label{sec:separation}

\begin{figure*}
            \centering
            \includegraphics[width=0.97\textwidth]{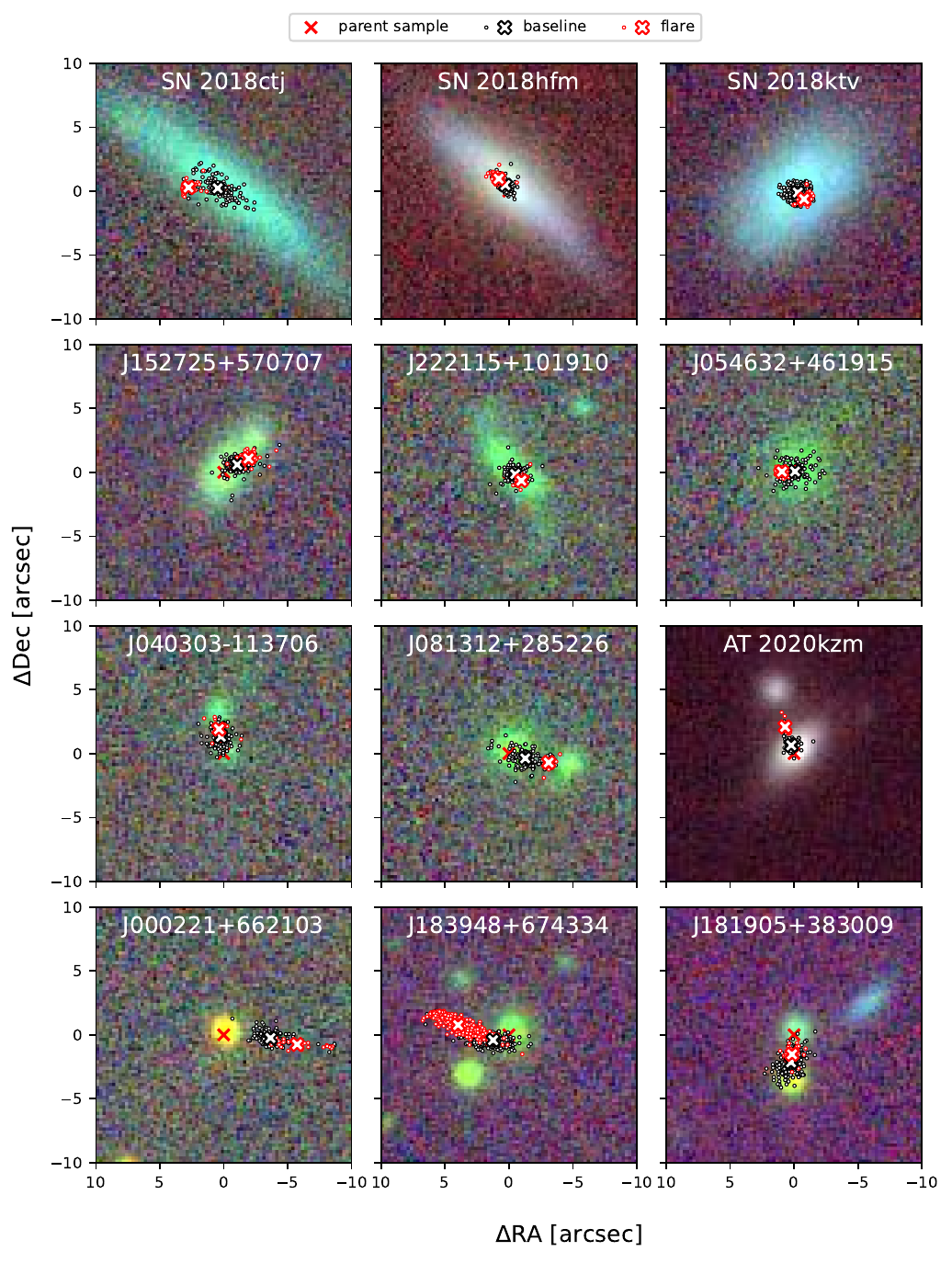}
            \caption{Pan-STARRS cutouts around the parent sample position of flares with a high angular offset. The red solid cross marks the position of the parent sample galaxy, the black and red circles the position of the single exposure photometry associated with the baseline and the flare, respectively, and the black and red filled cross the respective mean position.}
            \label{fig:offsets}
        \end{figure*}

        We calculated the median position of the single-exposure photometry data points for the baseline and the flare data and found 39 objects with a separation of more than \qty{0.5}{\arcsec}. This could indicate that the transient is not related to the SMBH in the nucleus of the host. \\
        Three offset flares are the confirmed supernovae discussed in Section \ref{sec:sn} and are shown in the top row of Figure \ref{fig:offsets}. The second row shows three more flares where the offset looks real which makes them promising supernova candidates. \\
        The first two panels of the last row show objects with an offset where the source of the offset can not be determined. There is no visible host galaxy at the position of the flare in the Pan-STARRS images. J183948+674334 is surrounded by three other sources which could cause the offset although the position of the flare is in between the parent sample galaxy and two of its neighbors. \\
        The remaining objects suffer from a confusion of host galaxies. The third row of Figure \ref{fig:offsets} shows the three objects where the baseline data is associated with the position given by the parent sample. Still, the flare could be associated with a neighboring galaxy. For the 28 remaining offset flares, the opposite is the case. The flare data is more consistent with the parent sample position and the baseline data is skewed towards a neighboring galaxy. An example is shown in the last panel. The confusion is probably due to a host that is too dim in the infrared to generate detections in the single-exposure photometry. The baseline, which is associated with the brighter neighbor is thus too bright causing an underestimation of the flare luminosity.

    \end{appendix}

\end{document}